\documentclass[twocolumn,amsmath,amssymb,floatfix]{revtex4}

\usepackage{graphicx}% Include figure files
\usepackage{dcolumn}% Align table columns on decimal point
\usepackage{bm}% bold math

\begin{document}

%\preprint{preprint-number}

\title{Study of first-order interface localization-delocalization transition \\ in thin Ising-films using Wang-Landau sampling}

\author{B. J. Schulz}
\email{schulzb@uni-mainz.de}
\author{K. Binder}
\affiliation{\mbox{Institut f\"ur Physik, WA331, Johannes Gutenberg Universit\"at, Staudinger Weg 7, D55099 Mainz, Germany}}
\author{M. M\"uller}
\affiliation{\mbox{Department of Physics, University of Wisconsin-Madison, 1150 University Avenue, Madison, WI 53706-1390}}
\date{\today}
\begin{abstract}
Using extensive Monte Carlo simulations, we study the interface localization-delocalization transition of a thin Ising film with 
antisymmetric competing walls 
for a set of parameters where the transition is strongly first-order. This is achieved by estimating the density of states 
(DOS) of the model by means of Wang-Landau sampling (WLS) in the space of energy, using both, single-spin-flip as well as 
N-fold way updates. From the DOS we calculate canonical averages related to the configurational 
energy, like the internal energy, the specific heat, as well as the free energy and the entropy. By sampling microcanonical 
averages during simulations we also compute thermodynamic quantities related to magnetization like the reduced fourth order
cumulant of the order parameter. We estimate the triple temperatures of infinitely large systems for three different film thicknesses via
finite size scaling of the positions of the maxima of the specific heat, the minima of the cumulant and the equal weight criterion 
for the energy probability distribution. The wetting temperature of the semi-infinite system is computed with help of the Young equation. 
In the limit of large film thicknesses the triple temperatures are seen to converge towards the 
wetting temperature of the corresponding semi-infinite Ising model in accordance with standard capillary wave theory. 
We discuss the slowing down of WLS in energy space as observed for the larger film thicknesses and lateral linear dimensions. In case of WLS 
in the space of total magnetization we find evidence that the slowing down is reduced and can be attributed to persisting free energy barriers 
due to shape transitions.
%
%
%Valid PACS numbers may be entered using the \verb+\pacs{#1}+ command.
\end{abstract}
%
%
%\pacs{05.70.Fh, 05.70.Np, 64.60.Cn, 02.70.Uu}% PACS, the Physics and Astronomy
                             % Classification Scheme.

%
%64.60.Fr,68.445.Gd,68.35.Rh
%

%\keywords{Ising model, Interface localization-delocalization, Monte Carlo, Wang-Landau sampling}%Use showkeys class option if keyword
%                              %display desired
%
\maketitle
%
%%%%%%%%%%%%%%%%%%%%%%%%%%%%%%%%%%%%%%%%%%%%%%%%%%%%%
%%%%%%%%%%%%%%%%%%%%%%%%%%%%%%%%%%%%%%%%%%%%%%%%%%%%%
%%%%%%%%%%%%%%%%%%%%%%%%%%%%%%%%%%%%%%%%%%%%%%%%%%%%%
%
\section{\label{introduction} Introduction}
\begin{figure}
\includegraphics[clip,width=0.46\textwidth]{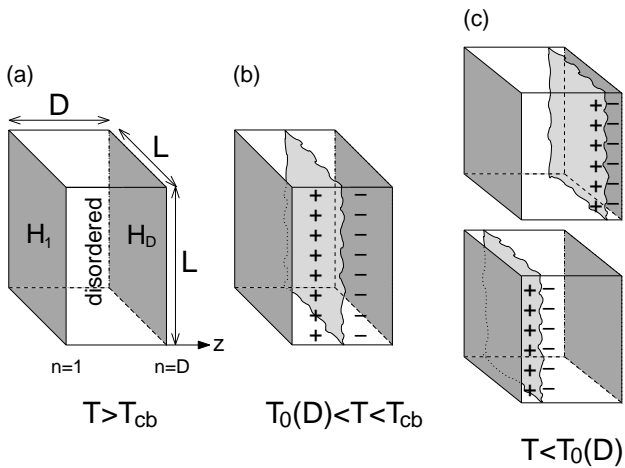}
\caption{(a) Thin film geometry with two free surfaces at $n=1$ and $n=D$ (shaded gray) on which magnetic surface fields
$H_1$ and $H_D$ act. Here, the surface at $n=1$ favors spin up (+), while the surface at $n=D$ favors spin down (-). 
Parallel to the $L\times L$ surfaces, periodic boundary conditions are imposed. 
(b) Delocalized Interface. (c) Interface located at either of the two surfaces.
%In the soft-mode phase the interface fluctuates 
%delocalized in the center of the film (b) while it becomes located at either of the two surfaces below  $T_{\mathrm{0}}(D)$ (c).
\label{schulz1}}
\end{figure}

%
%%%%%%%%%%%%%%%%%%%%%%%%%%%%%%%%%%%%%%%%%%%%%%%%%%%%%
%%%%%%%%%%%%%%%%%%%%%%%%%%%%%%%%%%%%%%%%%%%%%%%%%%%%%
%%%%%%%%%%%%%%%%%%%%%%%%%%%%%%%%%%%%%%%%%%%%%%%%%%%%%
%
The restriction of the geometry of a condensed-matter system has fundamental impact on a phase transition. In a finite system, 
sharp phase transitions can no longer occur, since the free energy is then an analytic function of its independent variables and
 the transition is rounded off and shifted. 
A particular realization of a confined geometry in $d=3$ dimensions, playing a pivotal role due to its fundamental importance 
in material science and technology, are thin films, infinitely extended in two directions but of finite thickness $D$, where the transition is now 
not only shifted away from its bulk value, corresponding to $D\rightarrow \infty$, but also changes its character from three-dimensional 
to two-dimensional.
As an example we may consider here a fluid near a gas-liquid coexistence in the bulk, or similarly, an ($A$,$B$) binary mixture or alloy near 
phase coexistence, 
confined between two parallel walls.\\
Of particular interest is the case, where the two walls of the system prefer different phases, i.e., one wall 
favors high-density liquid (or $A$-particles) while the other one prefers low-density gas (or $B$-particles), 
which is commonly termed ``competing walls''.
A generic model for such systems actually is the nearest neighbor Ising model in a thin film geometry where one now has
two surfaces a distance $D$ apart, on which magnetic surface fields $H_1=-H_D$ of
opposite sign but equal magnitude act in order to mimic the competing walls (cf. Fig.~\ref{schulz1}). In addition one
allows for a different interaction $J_{\mathrm{s}}>0$ between nearest neighbors located in the surfaces, while nearest neighbors in the 
bulk interact with a coupling $J>0$.
The meaning of the magnetic surface fields becomes apparent, when reinterpreting the Ising Hamiltonian as a lattice gas for a fluid or a mixture, 
where Ising spins $S_i=-1$ or $S_i=+1$ 
now correspond to lattice sites $i$ being empty or occupied, or being taken by an $A$-particle or a $B$-particle,
respectively. Then, surface magnetic fields translate into chemical potentials, i.e.,
binding energies to the walls. 
%
%%%%%%%%%%%%%%%%%%%%%%%%%%%%%%%%%%%%%% schulz1 %%%%%%%%%%%%%%%%%%%%%%%%%%%%%%%%%%%%%%%%%%%
%
%
%%%%%%%%%%%%%%%%%%%%%%%%%%%%%%%%%%%%%%%%%%%%%%%%%%%%%%%%%%%%%%%%%%%%%%%%%%%%%%%%%%%%%%%%%%%%
%
\\
Remarkably, the transition that one encounters in the Ising film differs from the transition in the bulk system at $T_{\mathrm{cb}}$
\cite{ParryEvansWetting,Swift,Indekeu,ParryReply,ParryEvans,BinderLandauFerrenberg,BinderLandauFerrenberg2,Ferrenberg}: 
For all finite thicknesses $D$ of the film, the transition at $T_{cb}$ is completely rounded off and no singular behavior shows up, despite
the fact that the system is infinite in the other directions. Instead, one observes a transition at a lower temperature 
$T_{\mathrm{0}}(D)<T_{\mathrm{cb}}$, at which the system changes from a state with a delocalized interface running parallel 
to the walls in the center of the film ($T_{\mathrm{0}}(D)<T<T_{\mathrm{cb}}$), 
to a twofold degenerate state ($T<T_{\mathrm{0}}(D)$), where the interface is now localized near one of the two walls (cf. Fig.~\ref{schulz1}).\\
Most interestingly, for $D\rightarrow \infty$, the transition temperature $T_{\mathrm{0}}(D)$ of the interface 
localization-delocalization does not converge towards 
the bulk critical temperature $T_{\mathrm{cb}}$, but towards the wetting temperature $T_{\mathrm{w}}(H_1)$ at which a macroscopically thick liquid layer 
(spins pointing upwards) wets the surface in the corresponding semi-infinite system.
Thus, the nature of the transition at finite $D$ is seen to depend on the nature of the wetting transition in the underlying semi-infinite 
system. Upon enhancing the interaction $J_{\mathrm{s}}$ of spins in the surfaces with respect to the bulk interaction $J$ one can tune 
the wetting transition and thus the interface transition for finite film thicknesses $D$ to be of first order \cite{Ferrenberg}, i.e.,
$T_{0}(D)\equiv T_{\mathrm{tr}}(D)$ is now a triple point where the three phases shown in \mbox{Fig.~\ref{schulz1}(b)--(c)} coexist. 
By reducing the film thickness one may then pass through a tricritical point where the order of the transition changes from first 
to second order \cite{Swift,Ferrenberg,MuellerAlbanoBinder}.\\
Our paper is arranged as follows: First, we briefly introduce the thin film Hamiltonian and give a description of the 
employed Wang-Landau sampling (WLS) which aims at sampling the density of states (DOS) directly. 
The slowing down of WLS for our model, encountered especially for large system sizes is discussed. 
With regard to these difficulties we then propose to split the DOS in a branch contributing to the ordered phase and one 
contributing to the disordered phase, which we normalize separately. We then present the thermodynamic quantities calculated from 
the DOS and compute the infinite lattice triple temperatures from the various finite size data. Finally, the wetting temperature 
of the semi-infinite system is determined via the Young equation and the convergence of the triple temperatures 
towards the wetting temperature for increasing film thickness is examined. We close with a brief discussion of our results.
%
%
%%%%%%%%%%%%%%%%%%%%%%%%%%%%%%%%%%%%%%%%%%%%%%%%%%%%%
%%%%%%%%%%%%%%%%%%%%%%%%%%%%%%%%%%%%%%%%%%%%%%%%%%%%%
%%%%%%%%%%%%%%%%%%%%%%%%%%%%%%%%%%%%%%%%%%%%%%%%%%%%%
%
\section{\label{modelandsim} Model and simulation method}
\begin{figure}[t]
\includegraphics[clip,width=0.46\textwidth]{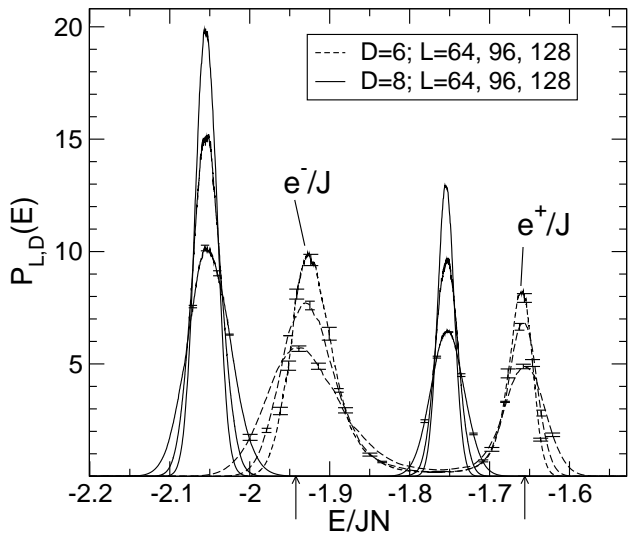}
\caption{Energy probability distributions $P_{L,D}$ at equal weight. The peak positions $e^-(L,D)$ and 
$e^+(L,D)$ (indicated for $D=6$ and $L=128$) 
define the finite volume latent heats $\Delta e(L,D)=e^+(L,D)-e^-(L,D)$. 
%
%The depicted probability distributions 
%were calculated from the density of states $g(E)$, as obtained from the individual WLS simulations, and were finally averaged. 
%
Arrows pointing on the energy axis indicate the interval $I_{\mathrm{center}}$, Eq.~(\ref{icenter}), 
%over which additional simulations have been performed 
in case of $L=128$ and $D=6$. \label{schulz2}}
\end{figure}

%
%%%%%%%%%%%%%%%%%%%%%%%%%%%%%%%%%%%%%%%%%%%%%%%%%%%%%
%%%%%%%%%%%%%%%%%%%%%%%%%%%%%%%%%%%%%%%%%%%%%%%%%%%%%
%%%%%%%%%%%%%%%%%%%%%%%%%%%%%%%%%%%%%%%%%%%%%%%%%%%%%
%
We consider the Ising Hamiltonian on a cubic lattice in a $L\times L \times D$ geometry (cf.~Fig.~\ref{schulz1}(a)), where 
$N=L^2D$ is the total number of spins $S_i$: 
\begin{eqnarray}
{\mathcal H} & = &-J\sum\limits_{\langle i,j\rangle_{\mathrm{b}}}S_i S_j
-J_s\sum\limits_{\langle i,j\rangle_{\mathrm{s}}} S_i S_j -H\sum\limits_{i}S_i \nonumber \\
& &-H_1\sum\limits_{i\in\mathrm{surface}\,1}S_i-H_D\sum\limits_{i\in\mathrm{surface}\,D}S_i. 
\label{thini}
\end{eqnarray}
Here, the sum $\langle i,j\rangle_{\mathrm{b}}$ runs over all pairs of nearest
neighbors where at least one spin is not located in one of the surfaces and the sum 
$\langle i,j\rangle_{\mathrm{s}}$ runs over all pairs of nearest neighbors with both spins located 
in one of the two surfaces. 
In this paper we study three different film thicknesses $D=6$, $8$, $12$, and linear lateral dimensions ranging from $L=16$ 
to $L=128$ (for the two largest choices of $D$ the minimal $L$ is $L=32$ and $L=48$, respectively). 
We restrict ourselves here to antisymmetric surface fields $H_1=-H_D$ and bulk field, $H=0$.
By virtue of the symmetry there is
an exact degeneracy of the phases where the interface is bound to either of the surfaces, and the triple
point and the phase coexistence below $T_0(D)$ occurs at $H=0$. We do not study pre-wetting like
phase coexistence for $T>T_0(D)$ and $H \neq 0$.
Specifically we choose $H_1/J=0.25$ and $J_{\mathrm{s}}/J=1.5$. 
For these parameters the interface localization-delocalization transition is clearly first-order for all thicknesses $D$. 
Already for a smaller surface-to-bulk coupling ratio $J_\mathrm{s}/J=1.45$, the transition turned out to be so
strongly first order according to the study of Ferrenberg et. al \cite{Ferrenberg}, that lattices with $D=8$ and $L>32$ could not be 
equilibrated using a standard canonical heatbath algorithm. The reason for such difficulties can be seen directly from the canonical 
probability distribution, $P_{L,D}(E)$, of the energy that develops two pronounced peaks at the transition point, corresponding to the 
coexisting ordered $(-)$ and disordered phases $(+)$ which are separated by a deep minimum  $P^{\mathrm{min}}_{L,D}(E)$ 
corresponding to the mixed phase configurations (cf. Fig.\ref{schulz2}). 
%
%
%%%%%%%%%%%%%%%%%%%%%%%%%%%%%%%%% schulz2 %%%%%%%%%%%%%%%%%%%%%%%%%%%%%%%%%%%%%%%%%%%%%%%%%
%
%
%%%%%%%%%%%%%%%%%%%%%%%%%%%%%%%%%%%%%%%%%%%%%%%%%%%%%%%%%%%%%%%%%%%%%%%%%%%%%%%%%%%%%%%%%%%%%%%%
%
%%%%%%%%%%%%%%%%%%%%%%%%%%%%%%%% schulz3 %%%%%%%%%%%%%%%%%%%%%%%%%%%%%%%%%%
%
\begin{figure*}[t]
\includegraphics[width=0.9\textwidth]{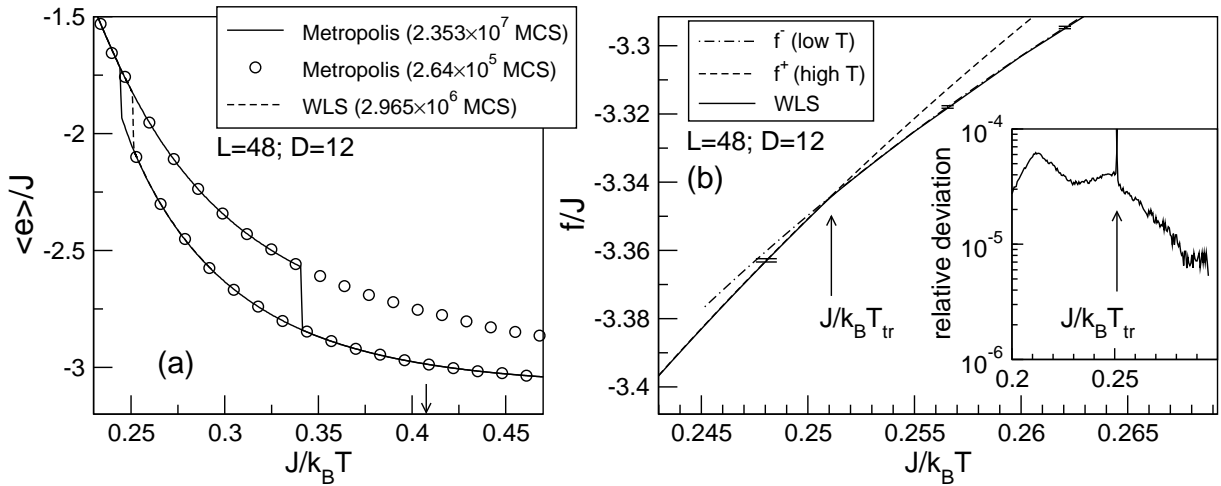}
\caption{(a) Energy hysteresis curves. 
%for a system of size $D=12$ and $L=48$. 
Cooling (heating) was performed at a rate of $J\Delta\beta/\mathrm{MCS}=4.3403 \cdot 10^{-6}$ 
(open circles, note that not all data points are plotted) and 
in steps of $J\Delta \beta=0.0005$, using $100$ MCS for equilibration at each $\beta$ and another $10^4$ MCS for measuring the 
energy (solid line). The equilibrium curve obtained from WLS in the space of energy is also shown.
The roughening temperature $J/k_{\mathrm{B}}T_{\mathrm{R}}=0.40758(1)$ \cite{HasenbuschPinnRough} is indicated by an arrow. 
%$0.00005 \le J\beta \le 1.16405$ 
(b) Low and high temperature branch of the free energy per site $f^{\pm}$ as obtained from thermodynamic integration, 
which yields $J/k_{\mathrm{B}}T_{\mathrm{tr}}(D=12)=0.2511(10)$. %The free energy as obtained from WLS is shown for comparison. 
The relative deviation 
\mbox{$|f_{\mathrm{WLS}}-f^{\pm}|/f_{\mathrm{WLS}}$} between the thermodynamic integration and the WLS result is plotted in the inset. 
\label{schulz3}}
\end{figure*}
%
%%%%%%%%%%%%%%%%%%%%%%%%%%%%%%%%%%%%%%%%%%%%%%%%%%%%%%%%%%%%%%%%%%%%%%%%%%%%
%
Here, one has additional interfaces in the system which cost an extra free energy 
$\Delta F_{L,D}=\gamma DL$, where $\gamma$ is of the order of the interface tension between the two oppositely oriented 
domains of spins. This yields $P^{\mathrm{min}}_{L,D}(E) \propto \exp(-\beta\Delta F_{L,D})$, where $\beta=1/k_{\mathrm{B}}T$ denotes
the inverse temperature. 
Hence, any simulation technique which aims at sampling a canonical energy probability distribution $\propto g(E)\exp(-E/k_{\mathrm{B}}T)$ 
directly will become trapped in the phase in which the system was initially prepared and may practically never escape from there, even 
in case of relatively small systems.\\
In order to give an example for the strong metastability, Fig.~\ref{schulz3} shows 
hysteresis-loops of the internal energy per spin $\langle e \rangle\equiv\langle E \rangle/N$ 
which were recorded using a conventional Metropolis Monte Carlo algorithm for a system of size $D=12$ and $L=48$.
The simulations were started in the disordered phase. In case cooling is performed too fast (open circles in Fig.~\ref{schulz3}) 
one reaches the roughening temperature $T_{\mathrm{R}}$ 
while still being in the disordered soft-mode phase, i.e., the interface becomes flat in the center of the film and it becomes 
impossible to reach the ordered phase upon further cooling. Using a much larger simulational effort ($\sim 10^7$ MCS) 
one obtains a closed loop -- although the observed hysteresis is still huge -- which clearly indicates a phase transition in the 
range $0.244<J\beta_{\mathrm{tr}}<0.341$. Locating the exact transition point in this way would however require an enormous 
simulational effort even for the moderate system size at hand. An improvement results from thermodynamic integration of 
the low- and high-temperature branches of the internal energy, which yields the free energy per site $f$ \cite{BinderZ,Liebmann}:
\begin{equation}
\beta f(\beta)=\beta_{\mathrm{ref}}f(\beta_{\mathrm{ref}})+\int_{\beta_{\mathrm{ref}}}^{\beta} \langle e \rangle_{\beta'}\mathrm{d}\beta'.
\label{TI}
\end{equation}
For the reference values we have regarded the spins as noninteracting at $J\beta_{\mathrm{ref}}=0.00005$, i.e., $f(\beta_{\mathrm{ref}})=
-\beta_{\mathrm{ref}}^{-1}\ln 2$, while 
on the low temperature side the free energy was matched with a series expansion based on the first two excited states at $J\beta_{\mathrm{ref}}
=1.10005$. 
The crossing point of both branches of the free energy then yields the transition point, which can be determined with an 
accuracy of $0.4\%$.\\
The result, that the correct location of the first order transition is not in the middle of the hysteresis loop but very close to its end at the 
high temperature side (dashed curve in Fig.~\ref{schulz3}) is very surprising at first sight. It should be noted however, that the hysteresis 
observed in Monte Carlo simulations has 
nothing to do with the ``Maxwell equal area rule'' of mean field theories, but is of kinetic origin. The almost free interface in the center of 
the film is very slowly relaxing and feels only a very weak potential from the walls, and thus is much more metastable than the state where the 
interface is tightly bound to one of the walls.
\subsection{Wang-Landau Sampling}
In order to avoid problems due to metastability and to further increase accuracy, 
we have decided to use Wang-Landau sampling (WLS) 
\cite{WangLandau1,WangLandau2,SchulzBinderMuellerLandau,SchulzBinderMueller} in order to compute thermodynamic 
quantities of the systems via estimating the density of states (DOS) of Hamiltonian (\ref{thini}).
In WLS one accepts trial configurations with probability $\min\left[1,g(E)/g(E')\right]$, where
$g(E)$ is the DOS and $E$ and $E'$ are the energies of the current and the proposed
configuration, respectively. At each spin flip trial the DOS is modified $g(E)\rightarrow g(E)\cdot f_i$ by
means of a modification factor $f_i$, which is reduced according to $f_{i+1}=\sqrt{f_i}$ in case the 
recorded energy histogram $H(E)$ is flat within some percentage $\epsilon$ of the average energy histogram, i.e., 
$H(E) \geq \epsilon\langle H(E')\rangle_{E'}$ for all $E$. $H(E)$ is then reset to zero and the procedure is repeated
until a flat $H(E)$ is achieved using a final modification factor $f_{\mathrm{final}}$. In practice one samples a logarithm 
of the DOS, i.e., $\log_{10} g(E)$ since $g(E)$ may become very large and modifying the DOS then corresponds to adding a small
modification increment $\Delta s_i= \log_{10}f_i$. The implementation of the single-spin-flip WLS is straightforward and we refer 
the reader to Refs. \cite{WangLandau1,WangLandau2,SchulzBinderMuellerLandau} for details. 
When considering systems with a large number of distinct energy levels it is useful to partition the entire energy range into 
adjacent subintervals in order to sample the DOS in a parallel fashion. 
For energy intervals that contain states with low degeneracy, e.g., the ground state, 
one can further accelerate WLS by combining it with the rejection-free N-fold way of Bortz et. al \cite{BortzKalosLebowitz,SchulzBinderMueller}. 
Here, the underlying idea is to partition all spins $S_\nu$, $\nu\in \{ 1,...,N\}$ into $M$ classes $c_\nu\in \{0,...,M-1\}$ 
according to the change in energy $\Delta E_{c_{\nu}}$ caused by flipping a spin $S_\nu$ at site $\nu$.
Making explicit use of $H_1=-H_D$ and $H=0$ in the Ising Hamiltonian (\ref{thini}),  we can evaluate $\Delta E_{c_{\nu}}$ as
follows:
\begin{eqnarray}
\!\!\Delta E_{c_{\nu}} &\!\! = \!\! & \left\{\begin{array}{r@{ }l} 2(Ju_{\nu}+J_{\mathrm{s}}v_{\nu} +H_1)S_{\nu}      
& \;\; \mbox{if}\;\; \nu \in \mbox{surfc.} 1
 \\ 2(Ju_{\nu}+J_{\mathrm{s}}v_{\nu}-H_1)S_{\nu} 
& \;\;\mbox{if}\;\; \nu \in \mbox{surfc.} D \\ 2 Ju_{\nu} S_{\nu} 
& \;\;\mbox{else}, \end{array}\right. 
\end{eqnarray}
where $S_\nu$ is the spin value before it is overturned and $u_\nu$ and $v_\nu$ denote sums over the nearest neighbor sites $\mu(\nu)$ of
site $\nu$:
\begin{eqnarray}
u_{\nu}& = & 
\left\{\begin{array}{c@{ }l}  \sum\limits_{\mu(\nu)}S_{\mu(\nu)}     
& \;\; \mbox{if}\;\; \nu \notin \mbox{surface}  \label{usum1}
 \\ \sum\limits_{\mu(\nu)\notin \mathrm{surface}}S_{\mu(\nu)}
& \;\;\mbox{if}\;\; \nu \in \mbox{surface}, \\ \end{array}\right.
\end{eqnarray}
and
\begin{eqnarray}
v_{\nu}&= & 
\left\{\begin{array}{c@{ }l}  0     
& \;\; \mbox{if}\;\; \nu \notin \mbox{surface} \label{vsum1} 
 \\ \sum\limits_{\mu(\nu) \in\mathrm{surface}}S_{\mu}
& \;\;\mbox{if}\;\; \nu \in \mbox{surface}, \\ \end{array}\right.
\end{eqnarray}
This results in a number of $M=27$ different classes. Within the context of N-fold way WLS the probability 
of any spin of a class $i$ being overturned is then given by
\begin{equation}
P(\Delta E_{i})=\frac{n(\mathcal{C},\Delta E_{i})}{N} p_{\mathcal{C}\rightarrow \mathcal{C}'},\quad i=1,...,M, 
\end{equation}
where $n(\mathcal{C},\Delta E_{i})$ denotes the number of spins of configuration $\mathcal{C}$ which belong to class $i$ 
and $p_{\mathcal{C}\rightarrow\mathcal{C}'}$ is given by
\begin{equation}
p_{\mathcal{C}\rightarrow\mathcal{C}'}=\left\{ \begin{array}{r@{\quad \mbox{if} \quad}l}
                               \min \left(1,\frac{g(E_{\mathcal{C}})}{g(E_{\mathcal{C}'})}\right) & E_{\mathcal{C}'}\in
                               I_{\mathrm{sub}} \\ 0 &
                               E_{\mathcal{C}'} \not\in I_{\mathrm{sub}}, \end{array} \right.
\end{equation}
where $I_{\mathrm{sub}}$ denotes the considered energy subinterval over which the DOS is sampled and 
$E_{\mathcal{C}'}=E_{\mathcal{C}}+\Delta E_i$. Classes are now chosen as follows. Firstly, one computes
the integrated probabilities for a spin flip within the first $m$
classes:
\begin{equation}
Q_{m}=\sum \limits_{i\leq m}P(\Delta E_{i}),\quad m=1,...,M\quad \mbox{and} \quad Q_{0}=0.
\label{q}
\end{equation}
By generating a random number $0<r<Q_{M}$ one then finds the class $m$ from which to flip a spin via 
the condition $Q_{m-1}<r<Q_{m}$. The spin to be overturned
is chosen from this class with equal probabilities, whereby $\log_{10} g(E)$ and the energy histogram 
are now updated by means of the average lifetime $\tau=1/Q_{M}$. A detailed description of the algorithm was 
given in Ref. \cite{SchulzBinderMueller}.\\
\subsection{Normalization of the DOS}
\begin{figure}[t] 
\includegraphics[clip,width=0.46\textwidth]{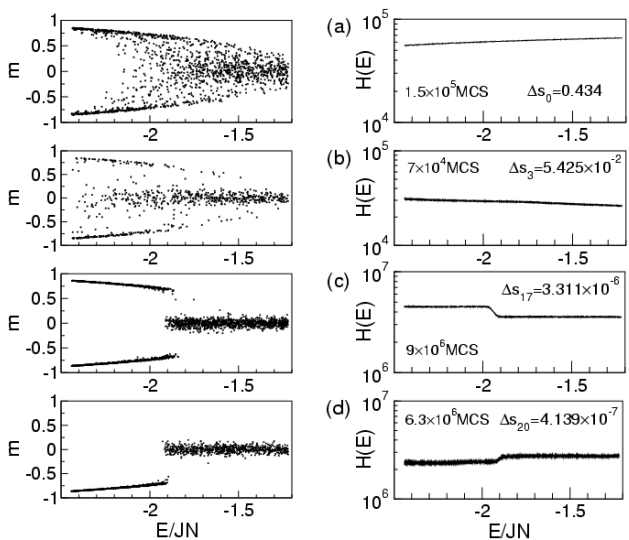}
\caption{Visited states ($E/JN$, $M/N$) (left hand side) and energy histogram H(E) (right hand side), recorded during WLS in the 
space of energy over the interval $E/JN \in [-2.4414,-1.2207]$ for different stages $i$ of the simulation. 
%where the modification increment $\Delta s_i$ is used to modify the DOS. 
The simulation used $3.844 \cdot 10^7$ MCS in total, with a flatness criterion for the energy histogram of $\epsilon=0.9$ and a 
final modification increment of $\Delta s_{\mathrm{final}}\approx4.139 \cdot 10^{-7}$. \label{schulz4}}
\end{figure}

In order to estimate the DOS using WLS, the considered energy range
\begin{equation}
E/JN \in I=\left[E_{\mathrm{ground}}/JN, 0.2\right], \label{range}
\end{equation}
where \mbox{$E_{\mathrm{ground}}=-[(3D-5)J+4J_{\mathrm{s}}]/JD$}, 
is the twofold degenerated ground state energy, was partitioned into several adjacent 
subintervals each containing an order of $10^2$ to $10^3$ distinct energy levels, which were sampled 
on a Cray T3E in a parallel fashion using mostly $64$ processors at a time. The DOS obtained from these simulations was then matched at the edges and 
suitably normalized, which we will describe in detail below.
For the system thicknesses $D=8$ and $D=12$, as well as for the largest choices of $L$ in case of $D=6$ ($L=96,128$) only one run was performed 
over the entire energy range (\ref{range}) denoted as basis-run, 
whereas all further runs have been restricted to a smaller energy range
\begin{equation}
E/JN \in I_{\mathrm{center}}=\left[E_1/JN,E_2/JN\right],  \label{icenter}
\end{equation}
covering the mixed phase region in between the peaks of the doubly peaked energy distribution. 
As an illustration, $I_{\mathrm{center}}$ is marked in Fig.~\ref{schulz8} by small arrows on the energy axis.
Thus, the entire energy range (\ref{range}) is decomposed as \mbox{$I=I_{\mathrm{left}}\cup  I_{\mathrm{center}} \cup I_{\mathrm{right}}$}, 
where we have $I_{\mathrm{left}}=[E_{\mathrm{ground}}/JN,E_1/JN]$ and $I_{\mathrm{right}}=[E_2/JN, 0.2]$.
Correspondingly, one obtains the density of states $g(E)$ by joining $g(E)$ 
estimated for the intervals $I_{\mathrm{left}}$ (taken from the basis run), $I_{\mathrm{center}}$, and $I_{\mathrm{right}}$ 
(again taken from basis run).\\
The single-spin-flip algorithm is more efficient in the regions covered by $I_{\mathrm{center}}$ which is due to the 
added expense of the N-fold way algorithm concerning the bookkeeping of classes. This was affirmed by a rough comparison between both 
implementations for $L=128$ and $D=12$. 
The flatness parameter $\epsilon$ varied between $0.8$--$0.95$, and the final modification 
increment was usually of order $\Delta s_{\mathrm{final}}\sim 10^{-9}$ which yielded an overall simulational effort of order $10^6$ MCS to $10^7$ MCS
for estimating the DOS over the range (\ref{range}).
As clear from the algorithm, WLS only yields a relative density of states, 
hence available reference values must be employed in order to get the absolute DOS $g(E)$.
Normalizing the simulational outcome firstly with respect to the twofold degeneracy of the ground state, i.e.,
the free energy $f$ will be exact for $\beta\rightarrow \infty$, 
it instructive to examine how this accuracy for low temperatures carries over to infinite temperature ($\beta\rightarrow0$), 
where the partition function $Z$ is dominated by the density of states around $E=0$, and one has 
$\lim_{\beta \rightarrow 0} \beta F(\beta )/N=-(1/N)\ln Z(\beta=0)=-\ln 2$.
Table \ref{freeenergylargeT} shows the latter quantity for all considered system sizes.
\begin{table}[t]
\begin{center}
{\small
\begin{tabular}{cccccc} \hline\hline
  $D$ & $L$ & $\#$ runs & $-\frac{\ln Z_{\beta=0}}{N}$ & $-\frac{\ln Z^{\mathrm{exact}}_{\beta=0}}{N}$ 
& $\frac{|\ln Z_{\beta=0}-\ln Z^{\mathrm{exact}}_{\beta=0}|}{|\ln Z^{\mathrm{exact}}_{\beta=0}|}$ \\ \hline
$6$ & $16$ & $6$ & $ -0.6932(5) $  & $ -0.693147 $  & $  0.0076 \%$ \\ 
$6$ & $24$ & $3$ & $ -0.6931(4) $  & $ -0.693147 $  & $   0.0061 \%$ \\  
$6$ & $32$ & $3$ & $ -0.6932(3) $  & $ -0.693147 $  & $   0.0082 \%$ \\  
$6$ & $48$ & $3$ & $ -0.69318(5) $  & $ -0.693147 $  & $  0.0048 \%$ \\  
$6$ & $64$ & $5$ & $ -0.69311(6) $  & $ -0.693147 $  & $  0.0049 \%$ \\  
$6$ & $96$ & $2$& $ -0.693112(2) $  & $ -0.693147$  & $  0.0051 \%$ \\  
%$6$ & $96$ & $3$ & $ -0.693103(2) $  & $ -0.693147 $  & $ 0.0063 \%$ \\  
$6$ & $128$ & $6$ & $ -0.693144(9) $  & $ -0.693147 $  & $ 0.00042 \%$ \\ \hline  
$8$ & $32$ & $4$ & $ -0.6930(2) $  & $ -0.693147 $  & $  0.027 \%$ \\  
$8$ & $48$ & $2$ & $ -0.69310(4) $  & $ -0.693147 $  & $  0.0075 \%$ \\  
$8$ & $64$ & $3$ & $ -0.69312(6) $  & $ -0.693147 $  & $  0.0038 \%$ \\  
$8$ & $96$ & $1$ & $ -0.69301 $  & $ -0.693147 $  & $  0.020 \%$ \\  
$8$ & $128$ & $1$  & $ -0.69304 $  & $ -0.693147 $  & $  0.016 \%$ \\   \hline
$12$ & $48$ & $3$ & $ -0.69240(2) $  & $ -0.693147 $  & $  0.107 \%$ \\  
$12$ & $64$ & $1$ & $ -0.692544 $  & $ -0.693147 $  & $  0.087 \%$ \\  
%$12$ & $96$ & $3$ & $ -0.692689(9) $  & $ -0.693147 $  & $  0.066 \%$ \\  
$12$ & $96$ & $2$ & $ -0.692686(3) $  & $ -0.693147 $  & $  0.067 \%$ \\  
$12$ & $128$ & $10$ & $ -0.69281(4) $  & $ -0.693147 $  & $  0.048 \%$ \\ \hline\hline
\end{tabular}}
\end{center}
\caption[$\ln Z(\beta=0)/N$ of a thin Ising film for different linear dimensions $L$ and $D$ (first order 
interface localization-delocalization transition, $J_{\mathrm{s}}/J=1.5$).]
{Logarithm of the partition function $-(1/N)\ln Z_{\beta=0}$ of a thin Ising film for different linear 
dimensions $L$ and $D$, in case the density of states is normalized with respect to the ground state. 
The value in brackets states the standard deviation. 
The exact value and the deviations from the latter are listed in the last two columns, respectively. 
For $L=32$ and $D=8$ the run showing the largest deviation from the exact value 
($-(1/N) \ln Z_{\beta=0}=-0.692624$) was excluded from data analysis. 
Then one has $ -(1/N) \ln Z_{\beta=0}=-0.69307(9)$. Under $\#$ runs we have listed the number of independent simulations. 
\label{freeenergylargeT}}
\end{table}
As can be seen from the table there is an increasing deviation from the exact value with increasing width of the film $D$. 
While the results for $D=6$ and $D=8$ (the latter for small sizes $L$) agree with the expected value, a deviation 
for the larger system sizes, especially $L=128$ and $D=12$ becomes apparent. 
This is related to a slowing down in the equilibration process in the multicanonical (Wang-Landau) ensemble for 
decreasing modification increment $\Delta s_i$, as illustrated in Fig.~\ref{schulz4}, which shows the visited
states ($E/JN$, $M/N$) and the energy histogram $H(E)$ recorded 
during Wang-Landau sampling for different stages $i$ of the simulation, where the modification 
increment $\Delta s_i$ is used to modify the density of states. In case one has a small number of tunneling events during a certain 
simulation stage $i$, $H(E)$ exhibits a kink at the barrier, since the stage is completed once the flatness criterion is fulfilled.
Correspondingly, $g(E)$ will suffer from large errors in the ordered phase, in case it is normalized 
using a reference in the disordered phase, and vice versa, errors will be enhanced in the disordered phase when using a reference in 
the ordered phase (ground state).\\
For $D=8$ (excluding $L=32$) and $D=12$ we therefore employed the following approach. Utilizing the fact that one has at least random walk 
behavior for small modification factor in each of the phases alone, we normalize the branch of the density of states $g_-(E)$ contributing 
to the low-energy ordered phase and the branch $g_+(E)$ contributing 
to the high-energy disordered phase separately, i.e., one has
\begin{eqnarray}
g(E) & = & \left\{\begin{array}{r@{ }l} g_-(E) & \;\; \mbox{for}\;\; E\le E_{\mathrm{cut}},
 \\ g_+(E) & \;\;\mbox{for}\;\; E> E_{\mathrm{cut}}, \end{array}\right. \label{shitsum}
\end{eqnarray}
where one obtains $g_-(E)$ by normalizing the simulational outcome $g(E)$ with respect to the ground state $g(E_{\mathrm{ground}})=2$ 
and is $g_+(E)$ obtained by normalizing $g(E)$ with respect to the total number of states 
\begin{equation}
\sum_E g(E) = 2^{L^2D}. \label{shitsum2}
\end{equation}
In Eq.~(\ref{shitsum}), $E_{\mathrm{cut}}$ is taken to be  
the energy for which the energy probability distribution, estimated directly from the simulational outcome $g(E)$, takes its minimum
in between the peaks at equal weight. Note, that in the sum 
$\sum_E g(E)=\sum_{E\le E_{\mathrm{cut}}} g_-(E)+\sum_{E>E_{\mathrm{cut}}} g_+(E)$ the term $\sum_{E\le E_{\mathrm{cut}}} g_-(E)$ is negligible.\\ 
The additional error which is introduced by this normalization procedure then depends on the contribution of the mixed phase configurations to the 
energy distribution (and the choice of $E_{\mathrm{cut}}$). However, since these mixed phase configurations are exponentially suppressed 
at the transition point, the error is expected to be of the same order, and correspondingly the error due to the choice of 
$E_{\mathrm{cut}}$ as well. Note, that already for $L=32$ and $D=8$ the double Gaussian approximation to the energy probability 
distribution, which neglects any mixed 
phase contribution, provides a reasonably good approximation to the measured distribution, apart from small deviations in 
in the tails (cf. Fig.~\ref{schulz8}).
%
%
%%%%%%%%%%%%%%%%%%%%%%%%%%%%%%%%%%%%%%%% schulz4 %%%%%%%%%%%%%%%%%%%%%%%%%%%%%%%%%%%%%%%%
%
%
%%%%%%%%%%%%%%%%%%%%%%%%%%%%%%%%%%%%%%%%%%%%%%%%%%%%%%%%%%%%%%%%%%%%%%%%%%%%%%%%%%%%%%
%
%
%
\subsection{Shape transitions}
\begin{figure}[t]
\includegraphics[clip,width=0.46\textwidth]{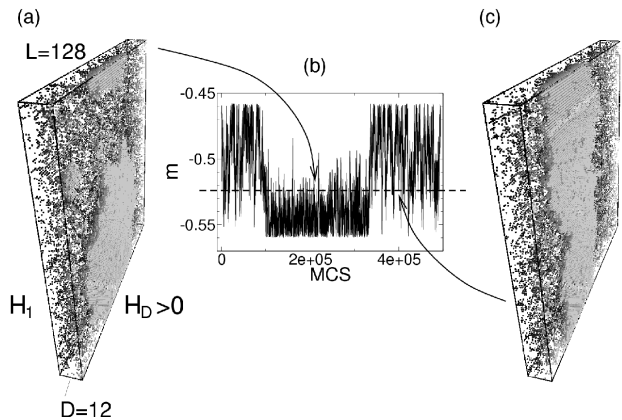}
\caption{(b) Selected part of a time-series of the total magnetization per spin $m=M/N$ as produced by WLS (single-spin-flip) in the
space of magnetization. (a) Droplet at surface $n=D$ where the positive field $H_{D}/J=0.25$ acts.
%
%, at total magnetization 
%per site $m=-0.523885$ indicated by the red horizontal line. 
%
(c) Percolated strip-like droplet. Note that in (a) and (c) only the positive spins are displayed as small spheres.
%, 
%which are grey-scale coded as function of depth, i.e., 
%
Those spins closest to the shown $L\times L$-surface are the lightest. \label{schulz5}}
\end{figure}

\begin{figure}[t]
\includegraphics[clip,width=0.46\textwidth]{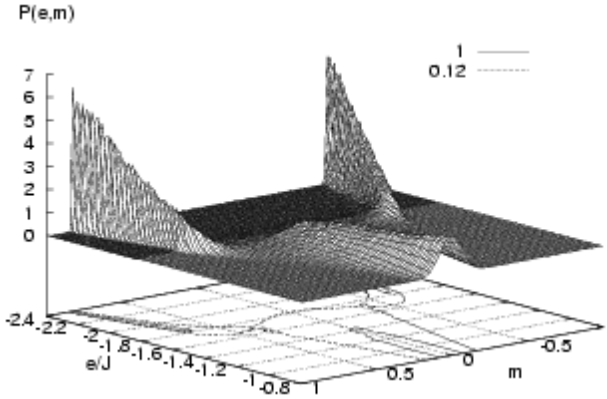}
\caption{Joint energy-order parameter distribution as obtained from WLS in the space of energy for a system size of $D=6$ and $L=16$. The 
distribution was recorded using a fixed DOS $g(E)$, which was taken from an usual adaptive WLS.\label{schulz6}}
\end{figure}

In Ref.~\cite{NeuhausHager}, Neuhaus and Hager addressed the severe problem of slowing down in simulations of first-order 
transitions in the multicanonical ensemble\cite{BergMulti}. 
This was exemplified by studying the two-dimensional Ising model 
on $L\times L$ square lattices (periodic boundary conditions) below the critical bulk temperature 
on the whole magnetization interval $[-L^2,L^2]$ whereby the sampling of configurations with 
magnetization $M=\sum_i S_{i}$ was biased with the inverse probability distribution of the magnetization $g^{-1}(M)$. 
Specifically, it was found in Ref. \cite{NeuhausHager} that these simulations suffered from a slowing down 
due to a discontinuous droplet-to-strip transition \cite{LeungZia}, i.e., $\tau \propto \exp(2 R\sigma L)$, where $\tau$ is the 
tunneling-time between droplet and strip configurations, $\sigma$ is the interfacial tension, and $R$ was measured to be 
$R=0.121(14)$. Note, that one has $R\approx 1$ for non-multicanonical simulations.\\
Of course, one needs a fairly good approximation to $g(M)$, in order to sample the considered Hamiltonian in the multicanonical ensemble. 
Within the framework of WLS one may therefore simulate the system 
at a certain inverse temperature $\beta$ of interest by 
employing the flipping probability (single-spin-flip Metropolis)
\begin{equation}
p_{\mathcal{C}\rightarrow\mathcal{C}'}  = \min \left[1,\frac{g(M_{\mathcal{C}})}{g(M_{\mathcal{C}'})} 
 \exp(-\beta \left[E_{\mathcal{C}'}-E_{\mathcal{C}}\right]) \right],
\end{equation}
for the transition from the state $\mathcal{C}$ to the state $\mathcal{C}'$. Each time a state with magnetization $M$ is visited, one updates
$g(M)$ according to $g(M) \rightarrow g(M)\cdot f_i$ in complete analogy to the case where $g(E)$ is used. 
Once this procedure has rendered $g(M)$ accurate enough, one makes a production run, where $g(M)$ is not altered anymore.
Thermodynamic quantities can then be obtained by reweighting to the canonical ensemble. \\
For the first order interface transitions in thin Ising-films as studied here, 
we have found evidence, that geometrical transitions in the ensemble realized by Wang-Landau sampling in the space of magnetization, 
indeed hamper the simulations. 
While this poses no problem for the smaller systems like $D=8$ and $L=32$ where WLS using $g(M)$ yields very good results (cf.~Fig.~\ref{schulz8}),
we observe pronounced effects for the largest considered system size. This is shown 
in Fig.~\ref{schulz5}(b) where part of a time series is depicted which was recorded for $D=12$ and $L=128$ 
during WLS 
%
%%%%%%%%%%%%%%%%%%%%%%%%%%%%%%%%%%%%%%%%%%%%%% schulz5 %%%%%%%%%%%%%%%%%%%%%%%%%%%%%%%%%%%%%%%%%%%%%%%%%
%
%
%%%%%%%%%%%%%%%%%%%%%%%%%%%%%%%%%%%%%%%%%%%%%%%%%%%%%%%%%%%%%%%%%%%%%%%%%%%%%%%%%%%%%%%%%%%%%%%%%%%%%%%%%%%%%%%%%%%%
%
sampling in the space of total magnetization. 
(Note also that the slowing down is more severe in case WLS using $g(E)$ is employed for the same system size as obvious from 
Fig.~\ref{schulz4}.) 
The simulation was restricted to the interval $m=M/N \in [-0.55949,-0.45776]$ after monitoring the time series of $m$ for a 
much larger interval $m\in [-0.91553,0.10173]$ where $\Delta s_{i}$ decreased from $5.0 \cdot 10^{-3}$ to $7.629 \cdot 10^{-8}$ 
over a 
simulation time of $1.632\cdot 10^{7}$ MCS at $J/k_{\mathrm{B}}T=0.249719$. 
The distribution $g(M)$ was then further iterated on the interval $m\in [-0.55949,-0.45776]$ where $\Delta s_{i}$ was refined  
from $1.0 \cdot 10^{-5}$ to $1.953 \cdot 10^{-8}$ within $7.27\cdot 10^{6}$ MCS and finally held fixed such that the depicted 
time series could be recorded. Configurations were thereby monitored along the estimated position of the barrier $m\approx-103000/N=-0.523885$.
Fig.~\ref{schulz5}(a) and (c) show snapshots of the two possible coexisting structures which are the
three-dimensional analogs (in the presence of a surface) to the droplet and strip shapes as studied in Ref. \cite{NeuhausHager}.
In case Wang-Landau sampling is performed in energy space, the governing mechanism of the slowing down is not determined by now. 
From the joint energy-order parameter distribution (Fig.~\ref{schulz6}) however,
%
%%%%%%%%%%%%%%%%%%%%%%%%%%%%%%%%%%%%%%% schulz6 %%%%%%%%%%%%%%%%%%%%%%%%%%%%%%%%%%%%%%%%%%%%%%%%
%
%
%%%%%%%%%%%%%%%%%%%%%%%%%%%%%%%%%%%%%%%%%%%%%%%%%%%%%%%%%%%%%%%%%%%%%%%%%%%%%%%%%%%%%%%%%
%
recorded for Wang-Landau sampling in the space of energy, one can at least conclude 
that one suffers from the fact that the ordered and the disordered phase are not distinctly separated in energy, as can be seen by 
inspecting the distribution of magnetization (along lines of constant energy) which shows a noticeable 
three-peak structure for a range of energies $e/J$. 
Further studies are clearly necessary in order to clarify whether there are connections to droplet related phenomena.\\
%
%%%%%%%%%%%%%%%%%%%%%%%%%%%%%%%%%% schulz7 %%%%%%%%%%%%%%%%%%%%%%%%%%%%%%%%%%%%%%%%%%%%%%%%%%%%%
%
\begin{figure}[t]
\includegraphics[clip,width=0.46\textwidth]{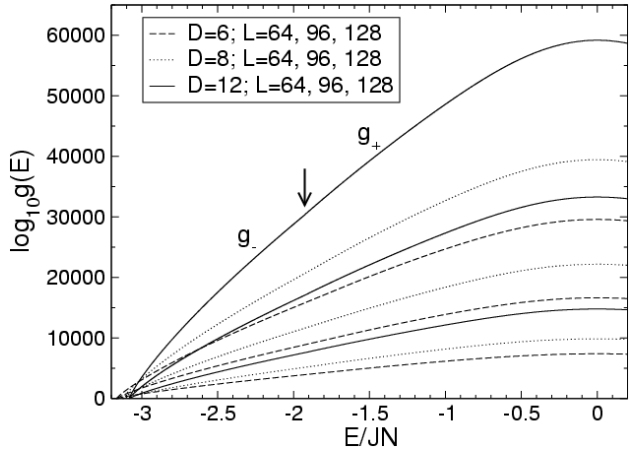}
\caption{Logarithm of the energy density of states $log_{10}(g(E))$ for thicknesses $D=6$, $8$, $12$ and linear dimensions 
$L=48,...,128$. Smaller choices for $L$ (in case of $D=6$ and $D=8$) are omitted in order to preserve clarity. Also indicated is 
the region where $E_{\mathrm{cut}}$, appearing in Eq.~(\ref{shitsum}), is typically located. Here, both branches of the density of states 
$g_-$ and $g_+$, are joined ($D=8,12$). In case of $D=6$, $g(E)$ was normalized solely with respect to the ground state degeneracy. 
\label{schulz7}}
\end{figure}
%
%%%%%%%%%%%%%%%%%%%%%%%%%%%%%%%%%%%%%%%%%%%%%%%%%%%%%%%%%%%%%%%%%%%%%%%%%%%%%%%%%%%%%%%%%
%
%
%%%%%%%%%%%%%%%%%%%%%%%%%%%%%%%%%%%%%%%%%%%%% schulz8 %%%%%%%%%%%%%%%%%%%%%%%%%%%%%%%%%%%%%%%%%%%%%%
%
\begin{figure*}[ht]
\includegraphics[width=0.9\textwidth]{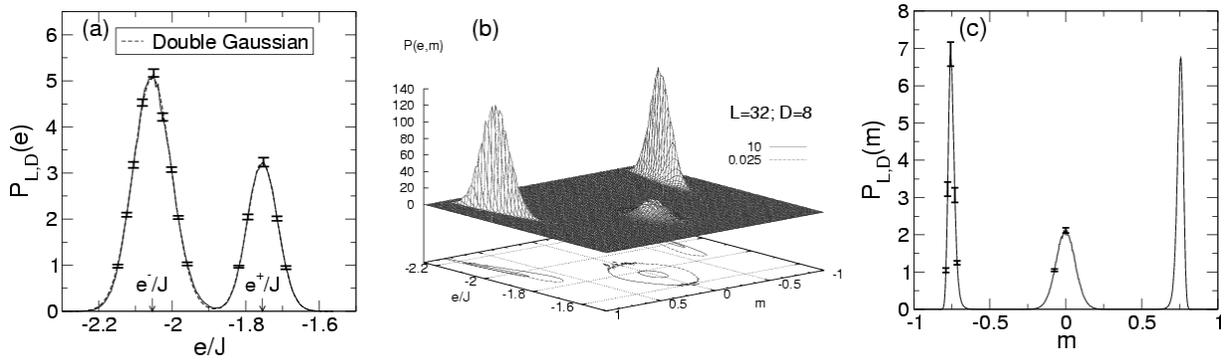}
\caption{Panel (a) shows the double Gaussian approximation (\ref{pld2}) to the energy probability distribution $P_{L,D}(e)$ for the system 
of size $L=32$ and $D=8$ at the finite volume transition point ($\beta_{\mathrm{tr}}(L,D)=0.247255(10)$) as obtained from WLS in the space of 
magnetization (single-spin-flip) by reweighting back to the canonical ensemble. Panel (b) shows the corresponding full joint energy- order 
parameter distribution $P_{L,D}(e,m)$, while (c) shows the projection onto the magnetization axis. \label{schulz8}}
\end{figure*}
%
%%%%%%%%%%%%%%%%%%%%%%%%%%%%%%%%%%%%%%%%%%%%%%%%%%%%%%%%%%%%%%%%%%%%%%%%%%%%%%%%%%%%%%%
%
%
%
%Most probably, 
%these difficulties could be avoided by considering WLS in the two-dimensional space of energy and
%order-parameter, but in general this approach is vain, since a two-dimensional density $g(E,M)$ can only be fitted into computer memory for very small
%systems, in spite of windowing both quantities, $E$ and $M$.\\
%
%
%
%%%%%%%%%%%%%%%%%%%%%%%%%%%%%%%%%%%%%%%%%%%%%%%%%%%
%%%%%%%%%%%%%%%%%%%%%%%%%%%%%%%%%%%%%%%%%%%%%%%%%%
%%%%%%%%%%%%%%%%%%%%%%%%%%%%%%%%%%%%%%%%%%%%%%%%
%
\section{\label{results} Simulation results}
%
%%%%%%%%%%%%%%%%%%%%%%%%%%%%%%%%%%%%%%%%%%%%%%%%%%
%%%%%%%%%%%%%%%%%%%%%%%%%%%%%%%%%%%%%%%%%%%%%%%%%%
%%%%%%%%%%%%%%%%%%%%%%%%%%%%%%%%%%%%%%%%%%%%%%%%%%%%%%
%
\subsection{Thermodynamic Quantities}
From the simulated DOS, as depicted in Fig.~\ref{schulz7}, we have calculated the first and second moment of the energy per spin 
\begin{equation}
\langle e^n \rangle=\frac{1}{NZ(\beta)} \sum\limits_{E}E^n g(E)\exp(-\beta E),
\label{internalenergy}
\end{equation}
and the specific heat
\begin{equation}
c=\frac{N}{k_{B}T^2}\left(\langle e^2\rangle-\langle e\rangle^{2}\right).
\label{specificheat}
\end{equation}
Furthermore, important quantities like the free energy per spin can be directly computed
\begin{equation}
f= - \frac{1}{N\beta}\ln Z(\beta)= - \frac{1}{N\beta}\ln \left[\sum\limits_{E}g(E)\mathrm{e}^{-\beta E} \right],
\label{gibbsfreeenergy}
\end{equation}
and the entropy per spin can be obtained from the internal energy (\ref{internalenergy}) and the 
free energy (\ref{gibbsfreeenergy})
\begin{equation}
s=\frac{\langle e \rangle-f}{T}.
\label{formulaentropy}
\end{equation}
%
%
%In order to compute canonical averages of the finite lattice order parameter
%
By measuring microcanonical averages $\langle \cdot \rangle_{E}$ during the last stage of a one-dimensional random walk in energy space, 
where $g(E)$ is updated with the smallest increment $\Delta s_{\mathrm{final}}$ we can also compute
canonical averages of the order parameter (and higher moments)
\begin{equation}
\left| m \right|=\frac{1}{N}\left|M\right|=\frac{1}{N}\left|\sum_{i=1}^{N} S_i \right|,
\end{equation}
i.e.,
\begin{equation}
\langle \left|m \right|^n \rangle  = \frac{\sum\limits_{E}\langle \left|m\right|^n \rangle_{E}\, 
g(E)\mathrm{e}^{-\beta E}}{\sum\limits_{E}g(E)\mathrm{e}^{-\beta E}}. \label{formulaabsm}
\end{equation}
Thus quantities like the finite lattice susceptibility $\chi$
\begin{equation}
\chi  =  \frac{N}{k_{\mathrm{B}}T}\left(\langle m^2 \rangle- \langle \left|m\right| \rangle^{2}   \right),\label{formulachi}
\end{equation}
as well as the fourth order cumulant $U_4$ on which we concentrate in the following and which is defined as
\begin{equation}
U_{4}=  1-\frac{\langle m^4 \rangle}{3\langle m^2 \rangle^{2}}, \label{u4}
\end{equation}
become accessible.\\
The distinctive feature of first-order phase transitions are phase coexistence and metastability. For the interface localization-delocalization 
transition considered here, this is reflected by jump discontinuities in the internal energy $\langle e \rangle$ as well as the (absolute) magnetization 
$\langle |m|\rangle$ per site, 
%
%%%%%%%%%%%%%%%%%%%%%%%%%%%%%% schulz9 %%%%%%%%%%%%%%%%%%%%%%%%%%%%%%%%%%%%%%%%%%%%%%%%%%%%%%%%%%%%%%%%%%%
%
\begin{figure*}[t]
\includegraphics[width=0.9\textwidth]{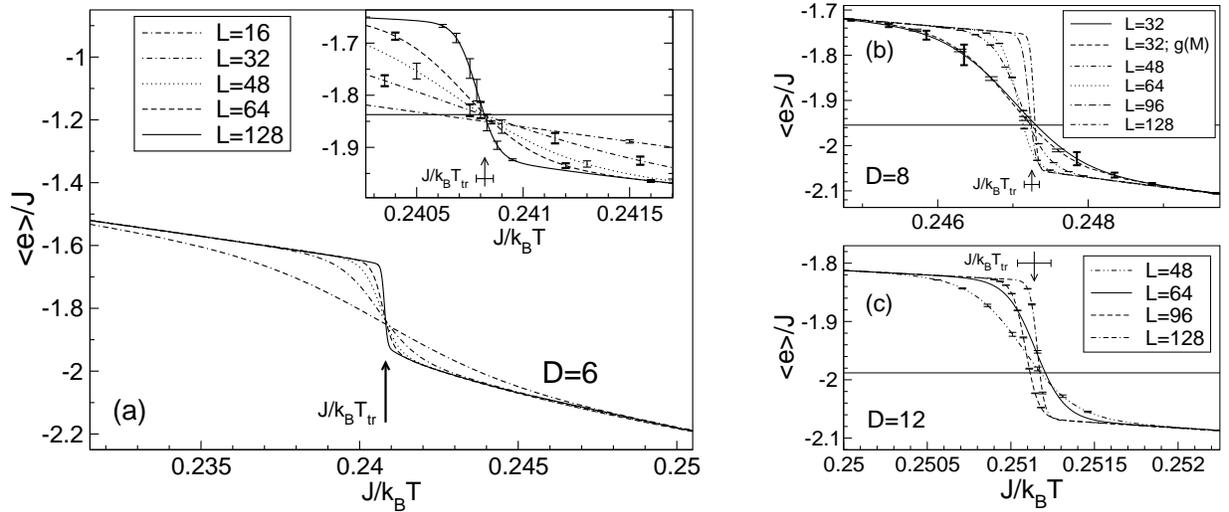}
\caption{Internal energy $\langle e \rangle$ for different linear dimensions $L$ and film thicknesses $D$. 
Estimates for the inverse temperature $J\beta_{\mathrm{tr}}(D)=J/k_{\mathrm{B}}T_{\mathrm{tr}}(D)$ of the triple point are indicated by arrows. 
%Also indicated is the latent heat $\Delta e = e^+-e^-$ in panel (c),
The horizontal solid lines mark the value $(e^++2e^-)/3$ where the curves are expected to cross.
%
%($e^+$ and $e^-$ were estimated from the 
%peaks of the $L=128$ energy probability distributions at equal weight of phases). 
%
In (b) the data obtained from WLS using $g(M)$ ($D=8$ and $L=32$) is plotted for comparison. 
Here, $g(E)$ for $D=8$ and $L=32$ was normalized solely with respect to the ground state. Within the inverse temperature range displayed in the 
inset of (a), the average relative errors in $\langle e \rangle$ for $D=6$ amount to $0.17\%$, $0.54\%$, $0.55\%$, 
$0.42\%$, and $0.45\%$ for $L=16,...,128$, respectively. For $D=8$ and $L=32$ the average error amounts to $0.18\%$ in the range 
$0.2455\le J/k_{\mathrm{B}}T \le 0.2485$, when using $g(M)$ while it is $1.0\%$ within the same range when using $g(E)$. Note, that for 
$D=8$ and $L=96$, $128$, as well as for $D=12$ and $L=64$ the DOS was determined only once, hence no error bars are displayed.
\label{schulz9}}
\end{figure*}
%
%%%%%%%%%%%%%%%%%%%%%%%%%%%%%%%%%%%%%%%%%%%%%%%%%%%%%%%%%%%%%%%%%%%%%%%%%%%%%%%%%%%%%%%%%%%%%%%%%%%%%%%%%%
%
%%%%%%%%%%%%%%%%%%%%%%%%%%%%%%%%%%%%%%%%%%%%% schulz10 %%%%%%%%%%%%%%%%%%%%%%%%%%%%%%%%%%%%%%%%%%%%%%%%%%%%%%%
%
\begin{figure}[t]
\includegraphics[clip, width=0.46\textwidth]{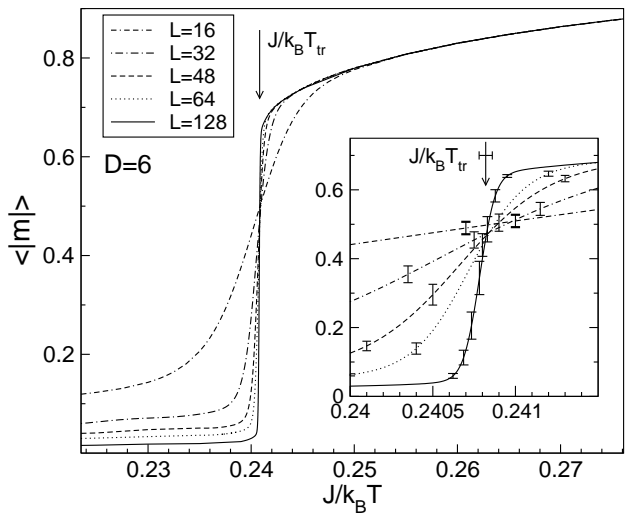}
\caption{Average absolute magnetization per spin $\langle |m|\rangle$ of a thin Ising film for different linear dimensions 
$L$ and film thicknesses $D$. 
%
%The infinite system transition point $J/k_{\mathrm{B}}T_{\mathrm{tr}}(D)$ is indicated by an arrow. 
%
Within the inverse temperature range displayed in the 
inset of (a), the average relative errors in $\langle |m| \rangle$ for $D=6$ amount to $1.0\%$, $5.2\%$, $6.3\%$, 
$6.1\%$, and $9.3\%$ for $L=16,...,128$, respectively. \label{schulz10}}
\end{figure}
%
%%%%%%%%%%%%%%%%%%%%%%%%%%%%%%%%%%%%%%%%%%%%%%%%%%%%%%%%%%%%%%%%%%%%%%%%%%%%%%%%%%%%%%%%%%%%%%%%%%%%%%%%%%%
%
%
%%%%%%%%%%%%%%%%%%%%%%%%%%%%%%%% schulz11 %%%%%%%%%%%%%%%%%%%%%%%%%%%%%%%%%%%%%%%%%%%%%%%%%%%%%%
%
\begin{figure*}
\includegraphics[clip, width=0.9\textwidth]{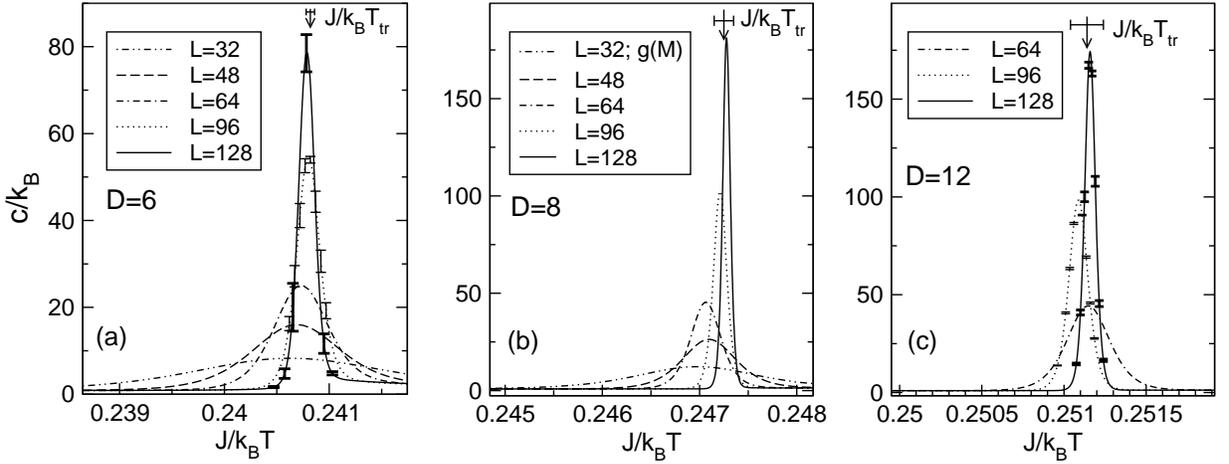}
\caption{Specific heat $c$ of a thin Ising film for different linear dimensions $L$ and film thicknesses $D$. 
%The inset of panel (a) shows the specific heat for $L=16$, $24$, and $32$ on a finer scale. 
%Estimates for the inverse temperature $J/k_{\mathrm{B}}T_{\mathrm{tr}}(D)$ of the triple 
%point are indicated by arrows. 
In the interval $0.2400 \le J/k_{\mathrm{B}}T \le 0.2415$ the average relative error for $D=6$ amounts to
$3.9\%$, $9.6\%$, $12.0\%$, $5.4\%$, and $15.9\%$ for $L=32,...,128$, respectively.
For $D=8$ and $L=32$, Wang-Landau sampling in $g(M)$ yields an average error of $2.6\%$ within the range $0.2455 \le J/k_{\mathrm{B}}T \le 0.2480$.
Note, that we have no statistics for $D=8$ and $L=96,128$ (b), as well as for $D=12$ in case of $L=64$ (c). 
\label{schulz11}}
\end{figure*}
%
%%%%%%%%%%%%%%%%%%%%%%%%%%%%%%%%%%%%%%%%%%%%%%%%%%%%%%%%%%%%%%%%%%%%%%%%%%%%%%%%%%%%%%%%%%%%%%%%
%
%
which are depicted in Figs.~\ref{schulz9} and \ref{schulz10}, respectively, and also by hysteresis effects encountered when heating and cooling 
the system as exemplified in Fig.~\ref{schulz3}. 
Considering the internal energy (Fig.~\ref{schulz9}) 
for fixed $D$ and varying linear dimension $L$, one can clearly see that one actually does not observe 
discontinuous jumps of the quantities in question but a continuous behavior that sharpens to the asserted step-like behavior with
increasing linear system size $L$. This rounding is related to the fact that a true phase transition can only occur in the thermodynamic limit, 
where in equilibrium, 
approaching the transition temperature from above, the energy of the system discontinuously jumps from $e^+$ (interface in the center of the film) 
to $e^-$ (interface tightly bound to the wall), while for a finite volume the system may jump back and forth 
between the latter states and the observed equilibrium behavior is thus continuous in temperature. 
The rounding of the transition in finite systems 
can also be observed for the specific heat $c$ depicted in Fig.~\ref{schulz11} 
which exhibits narrow peaks that are remnants of the $\delta$-function singularities one would get 
when differentiating the discontinuous energy in the infinite volume limit. Apart from the finite size rounding, one 
can see that the positions of the maxima of $c$ and the minimum of the fourth order cumulant $U_4$ (Fig.~\ref{schulz12}) 
are systematically shifted towards higher $\beta$-values for increasing linear dimension $L$. \\
From the crossings of the energy curves for different linear dimensions $L$, one can get a first idea about the achieved accuracy for the 
different film thicknesses $D$, because they should cross to a very good approximation in the point \cite{Borgs} 
\begin{equation}
(\beta_{\mathrm{tr}}(D),(e^++2e^-)/3), \label{crossingpoint}
\end{equation}
where $\beta_{\mathrm{tr}}(D)$ is the infinite system transition point. 
Hence, the crossing points for different $L$ 
\begin{equation}
\langle e(\beta_{\mathrm{cross}},L,D) \rangle=\langle e(\beta_{\mathrm{cross}},L',D) \rangle, \label{crossingpoint2}
\end{equation}
actually provide an estimator for the infinite system transition temperature, which is 
expected to deviate from $\beta_{\mathrm{tr}}(D)$ only by an amount exponentially small in system size \cite{Borgs}. 
As can be seen from the inset of Fig.~\ref{schulz9}(a) in case of $D=6$, the various crossings 
are indeed scattered in a narrow region around the extrapolated infinite volume transition point for $L\ge32$. For smaller values 
of $L$ exponential corrections still make a noticeable contribution.
For the larger thicknesses $D\ge 8$ the region where the energy curves cross is noticeably larger. Particularly, one 
obtains that the errors resulting from averaging over different runs are too small to fully account for the deviations 
(excluding $L=32$ for which $g(M)$ was employed). This is related to the fact that for the thicknesses $D=8$ and $D=12$ only a single run 
was performed over the entire energy range (\ref{range}) while further runs were restricted to the mixed phase region in between the peaks, 
because the slowing down, as described 
in the preceding subsection, was not foreseen. When one uses the normalization condition (\ref{shitsum}), the proper 
strategy would certainly be to enhance the simulational effort in the pure phases, down to the ground state and up to $E=0$, 
since the reference density of states is known for $T=0$ and $\beta=0$. This is necessary, in order to minimize the
accumulation of errors in the density of states, since the Wang-Landau method and
 similar adaptive algorithms, do in general not exhibit an error distribution 
that is flat in energy \footnote{Recently, an adaptive algorithm was proposed \cite{Tro} which aims at maximizing the number of round trips between 
both edges of an energy interval which has the additional benefit of exhibiting a flat error distribution.}.
Hence, for $D=8$ (excluding the simulation using $g(M)$) and $D=12$, we believe the true errors to be larger 
than the error bars displayed in Figs.~\ref{schulz9}(b)-(c), \ref{schulz12}(b), \ref{schulz11}(b)-(c) 
and when quantitatively referring to errors of the thermodynamic quantities, 
we thus restrict ourselves here to $D=6$, where we have reliable error estimates.\\ 
Exponential corrections to the crossing points are presumably much smaller than the scatter in the energy crossings for $D\ge8$ 
%(see also $\beta_{\mathrm{w}}$ as listed in Table \ref{tableequalweights}) 
and one may therefore conclude that the deviations 
in the crossings for $D\ge8$ are not due to corrections to scaling, but reveal the actual error in the density of states for this region. 
This is also the case for the other quantities like the specific heat for example (Fig.~\ref{schulz11}(b)-(c)). 
Thus, the analysis of the systems with larger thicknesses $D=8$ and $D=12$ is certainly more difficult and less accurate. \\
One can however roughly estimate the order of magnitude of the latter uncontrolled error, which also serves to support the above picture. 
For example, from the density of states of the largest system ($D=12$ and $L=128$), one can estimate, that a relative error in the 
density of states $g(E)$ of the order $\sim10^{-1}$ (referring to the results for the $50\times50$ $2D$ Ising model in Ref. \cite{SchulzBinderMueller} 
this seems to be a reasonable assumption), in the narrow region corresponding to the peak of the ordered phase 
of the energy probability distribution, can result in a displacement $\Delta \beta$ of the peak position 
$\beta_{c^\mathrm{max}}$ of the specific heat 
%(cf. Table \ref{tableequalweights}), 
and also of the step location of the internal energy, which is approximately of the order $\Delta \beta/\beta_{c^\mathrm{max}}\sim10^{-4}$. 
In case of $D=12$ and $L=48$, a relative deviation of this order could already be caused by a relative error in $g(E)$ which is of 
the order $\sim10^{-2}$ in the above region. These considerations comply well with the observed scatter. 
%
%
%
%%%%%%%%%%%%%%%%%%%%%%%%%%%%%%%%%%%%%%%%%%%%%%%%%%%%%
%%%%%%%%%%%%%%%%%%%%%%%%%%%%%%%%%%%%%%%%%%%%%%%%%%%%%%
%%%%%%%%%%%%%%%%%%%%%%%%%%%%%%%%%%%%%%%%%%%%%%%%%%%%%%
%
\subsection{\label{fss} Finite size scaling}
%
%
%%%%%%%%%%%%%%%%%%%%%%%%%%%%%%%%%%%%%%%%%%%%%%%%%%%%%%
%%%%%%%%%%%%%%%%%%%%%%%%%%%%%%%%%%%%%%%%%%%%%%%%%%%%%
%%%%%%%%%%%%%%%%%%%%%%%%%%%%%%%%%%%%%%%%%%%%%%%%%%%%%%%
%
%
%%%%%%%%%%%%%%%%%%%%%%%%%%%%%%%%% schulz12 %%%%%%%%%%%%%%%%%%%%%%%%%%%%%%%%%%%%%%%%%%%%%%
%
\begin{figure*}[t]
\includegraphics[width=0.9\textwidth]{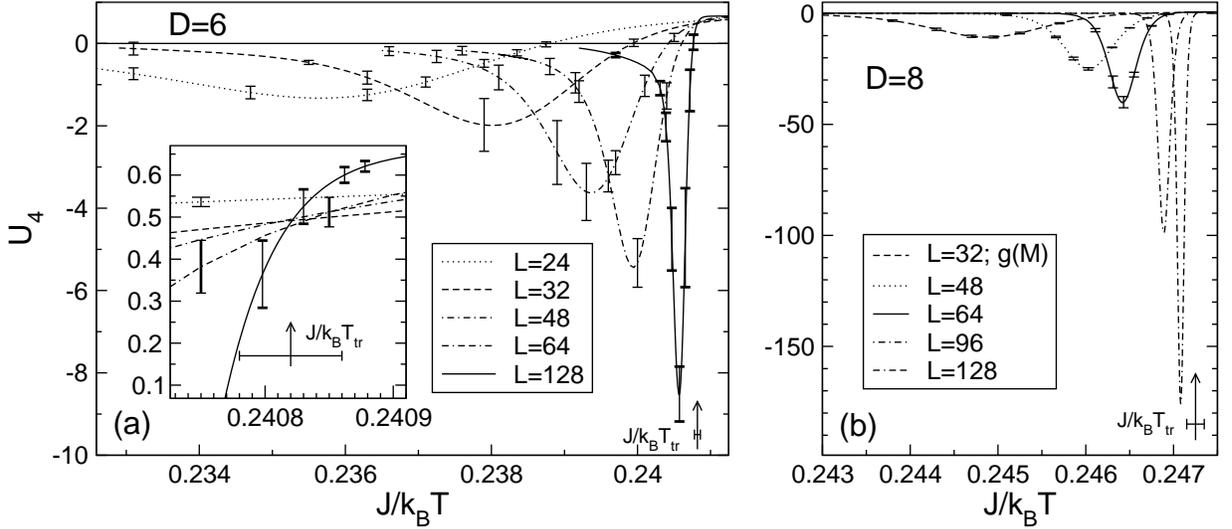}
\caption{Reduced fourth order cumulant of a thin Ising film for different linear dimensions $L$ and film thicknesses $D$. 
The inset of panel (a) shows the region where the cumulants for the various linear sizes $L$ cross ($D=6$). 
In the vicinity of the minima positions, the relative errors in $U_4$ in case of $D=6$ amount to
$11\%$, $30\%$, $17\%$, $11\%$, and $7\%$ for $L=24,...,128$, respectively.
WLS in the space of total magnetization $M$ with fixed $g(M)$ yields an error of $4\%$ ($D=8$, $L=12$).
Note, that for the data corresponding to $D=8$, as plotted in (b), we have no statistics for $L=96$ and $L=128$, i.e., for the latter 
sizes the DOS was estimated only once.
\label{schulz12}}
\end{figure*}
\begin{figure*}[t]
\includegraphics[width=0.9\textwidth]{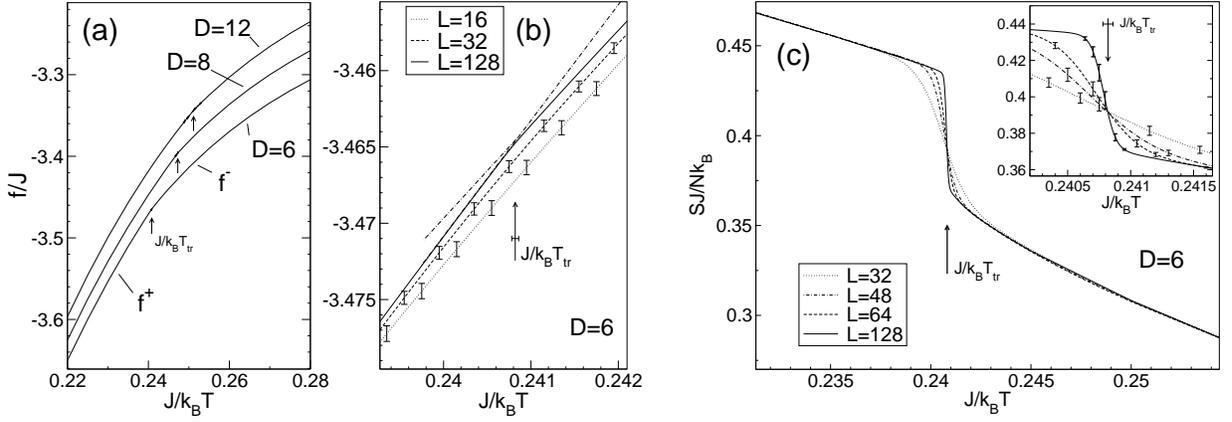}
\caption{(a) Free energy per spin $f$ of a thin Ising film for different linear dimensions $L$ and film thicknesses $D$. 
Note, that in (a) only the data for $L=128$ is plotted, while (b) shows $f$ on a finer scale. 
In the range $0.23\le J/k_{\mathrm{B}}T \le 0.27$, the average relative error in the free energy ($D=6$) is $0.0143\%$, $0.011\%$, 
and $0.00026\%$ for $L=16$, $32$, and $128$, respectively. (c) entropy per spin $s$ for $D=6$. The error in $s$ amounts to 
$0.64\%$, $0.64\%$, $0.47\%$, and $0.47\%$, within the range depicted in the inset of panel (c). 
%Estimates for the inverse 
%temperature $J/k_{\mathrm{B}}T_{\mathrm{tr}}(D)$ of the triple point are indicated by arrows.
\label{schulz13}}
\end{figure*}

%
%%%%%%%%%%%%%%%%%%%%%%%%%%%%%%%%%%%%%%%%%%%%%%%%%%%%%%%%%%%%%%%%%%%%%%%%%%%%%%%%%%%%%%%%%%%%%%%%
%
%
When one deals with second order phase transitions, the characteristic 
feature is a divergent spatial correlation length $\xi$ at the transition point
$\beta_{\mathrm{c}}$ (where one observes fluctuations on all length scales) implying power-law singularities in thermodynamic functions such as
the correlation 
length, magnetization, specific heat and susceptibility. 
This is in sharp contrast to a first order transition where the correlation length in the coexisting pure phases remains finite 
and concerning finite size scaling the volume of the system turns out to be the relevant quantity. 
For a thin film geometry where one has fixed $D$ and varying linear dimension $L$, finite size scaling 
will thus involve the quantity $L^2$. This can be shown by approximating the energy distribution $P_{L,D}(e)$ of the pure phases  
by a Gaussian \cite{Challa,Borgs,BorgsKotecky} centered around the infinite-lattice energy per spin 
$\langle e \rangle$ 
\begin{equation}
P_{L,D}(e)=\sqrt{\frac{L^2D}{2\pi k_{\mathrm{B}}T^2 c}} \exp\left[\frac{(e - \langle e\rangle)^2}{2k_{\mathrm{B}}T^2c}L^2D\right], \label{gaussianapprox}
\end{equation}
where $c$ denotes the infinite-lattice specific heat. 
Since one has phase coexistence at a first-order transition, the probability distribution of the energy 
will be double peaked at the transition point $\beta_{\mathrm{tr}}(D)=1/k_{\mathrm{B}}T_{\mathrm{tr}}(D)$, where $\langle e \rangle$  
jumps from $e^-$ (low energy phases, interface at one of the two walls) to $e^+$ 
(single high energy phase, interface centered in the middle of the film), i.e., the free energy branches $f^{\pm}$ intersect at a 
finite angle in the infinite system, as can be seen from Fig.~\ref{schulz13}(a), when inspecting the curves around the transition point (cf.~also Fig~3(b))
It is essentially this non-analyticity in the free energy, which gives rise to the discontinuous behavior of the internal energy. In a finite 
system however, the free energy remains differentiable and the intersection is rounded.\\
Hence, at the transition point, $P_{L,D}(e)$ is a superposition of two Gaussians (\ref{gaussianapprox}) centered at $\langle e \rangle=e^\pm$, 
while slightly away from the transition at $T=T_{\mathrm{tr}}+\Delta T$ they are centered at energies $e^\pm+c^\pm \Delta T$, where $c^\pm$
are the specific heats in the disordered (+) and ordered phases (-), which are assumed to be constant in the vicinity of the transition, i.e.,
for sufficiently small $\Delta T$. Each of the Gaussians is then weighted by Boltzmann factors of the corresponding free energies $f^\pm$, 
and one thus arrives at
\begin{widetext}
\begin{equation}
P_{L,D}(e)=A \left[ \frac{a^+}{\sqrt{c^+}} \exp\left[- \frac{[e - (e^++c^+\Delta T)]^2}{2k_{\mathrm{B}}T^2c^+} L^2D\right] 
           +\frac{a^-}{\sqrt{c^-}} \exp\left[- \frac{[e - (e^-+c^-\Delta T)]^2}{2k_{\mathrm{B}}T^2c^-} L^2D\right] \right],\label{pld2}
\end{equation}
\end{widetext}
where the weights $a^\pm$ are given by 
\begin{equation}
a^\pm=q^\pm\exp\left[\mp \frac{f^+ - f^-}{2k_{\mathrm{B}}T}L^2D\right], \label{aweights}
\end{equation}
and $A$ reads
\begin{equation}
A=\exp\left[ -\frac{(f^++f^-)}{2k_{\mathrm{B}}T} L^2D \right] \sqrt{\frac{L^2D}{2\pi k_{\mathrm{B}}T^2}}.
\end{equation}
Since we have a single high energy phase and two low energy ordered phases we set $q^+=1$ and $q^-\equiv q=2$ in the following. 
At the transition all phases have equal weight \cite{BorgsImbrie,Borgs} such that the area under the peak at $e^-$ is $q$ times the area under
the peak at $e^+$ 
which is satisfied by Eq.~(\ref{pld2}). 
Within the framework of the Ansatz (\ref{pld2}) one then proceeds by calculating the energy moments as usual via \cite{Challa}
\begin{equation}
\langle e^n \rangle = \frac{\int \mathrm{d}e' e'^n P_{L,D}(e')}{\int \mathrm{d}e' P_{L,D}(e') }. \label{mom}
\end{equation}
Computing then $\langle e \rangle$ at the transition point by means of Eq.~(\ref{mom}) we obtain
\begin{equation}
\langle e \rangle= \frac{e^++qe^-}{1+q}, \label{energyattrans}
\end{equation}
(horizontal lines in Fig.~\ref{schulz9}), which is exact, apart from exponential corrections due to mixed phase 
contributions which are neglected in the double 
Gaussian approximation. Upon using the fluctuation relation (\ref{specificheat}) or \mbox{$c=\mathrm{d} \langle e \rangle/\mathrm{d}T$} 
in conjunction with Eq.~(\ref{mom})
%
%
%%%%%%%%%%%%%%%%%%%%%%%%%%%%%%%%%%%%%%%%%%%%%%%%%%%%%%%% schulz13 %%%%%%%%%%%%%%%%%%%%%%%%%%%%%%%%%%%%%%%%%
%
%
%%%%%%%%%%%%%%%%%%%%%%%%%%%%%%%%%%%%%%%%%%%%%%%%%%%%%%%%%%%%%%%%%%%%%%%%%%%%%%%%%%%%%%%%%%%%%%%%%%%%%%%%%%%%%%%%
%
one can calculate the specific heat to leading order
\begin{eqnarray}
c & = & \frac{a^+c^++a^-c^-}{a^++a^-} \nonumber \\
  & & + \frac{[e^+ - e^-+(c^+ - c^-)\Delta T ]^2}{(a^++a^-)^2} \frac{a^+a^-}{k_{\mathrm{B}}T^2}L^2D, \label{approxspec}
\end{eqnarray}
which is seen to take its maximum for $a^+=a^-$ in Eq.~(\ref{pld2}). The position of the latter is thereby shifted away from 
the infinite lattices transition temperature by an amount of
\begin{equation}
\Delta T = T_{c^{\mathrm{max}}}(D,L) - T_{\mathrm{tr}}(D)= k_{\mathrm{B}}T_{\mathrm{tr}}^2\frac{\ln q}{\Delta e D} \frac{1}{L^2},\label{shiftTC}
\end{equation}
and the height of the peak is found to be
\begin{equation}
c^{\mathrm{max}}=\frac{c^++c^-}{2}+\frac{\Delta e^2 D}{4k_{\mathrm{B}}T_{\mathrm{tr}}^2}L^2, \label{specmax}
\end{equation}
where $\Delta e\equiv e^+-e^-$ is the latent heat.
For convenience we may re-express Eq.~(\ref{shiftTC}) in terms of the inverse temperature $\beta=1/k_{\mathrm{B}}T$ which yields
%
%
%\begin{equation}
%\Delta \beta\equiv\beta_{c^{\mathrm{max}}}(D)-\beta_{\mathrm{tr}}(D)=-\frac{\ln 2}{\Delta e} \frac{1}{L^2D}. \label{betashift}
%\end{equation}
%
%
\begin{equation}
\beta_{c^{\mathrm{max}}}(D,L)=\beta_{\mathrm{tr}}(D)-\frac{\ln 2}{\Delta e D}\frac{1}{L^2}. \label{betashift}
\end{equation}
Thus, the inverse temperature $\beta_{c^{\mathrm{max}}}(D)$ at which the specific heat peaks, provides 
a definition for a finite-lattice (pseudo) transition temperature from which the infinite-lattice transition temperature can be estimated via 
finite size scaling, i.e., by extrapolating $L\rightarrow \infty$.\\
%
%
%%%%%%%%%%%%%%%%%%%%%%%%%%%%%%%%%%%%%%%%%%%%%%%%%%%% schulz14 %%%%%%%%%%%%%%%%%%%%%%%%%%%%%%%%%%%%%%%%%%%%%%%%%%%
%
\begin{figure*}[t]
\includegraphics[clip, width=0.9\textwidth]{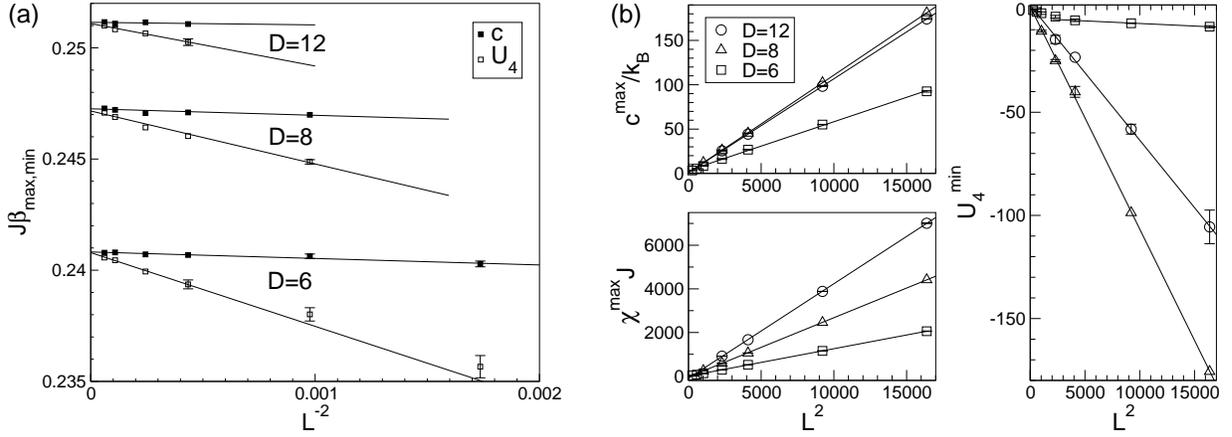}
\caption{Panel (a): extrapolation of peak positions $\beta_{\mathrm{max,min}}(D,L)$ of the specific heat $c^{\mathrm{max}}$ and 
the fourth order cumulant $U_4^{\mathrm{min}}$ for the different film thicknesses $D$. 
Panel (b): maxima of specific heat $c^{\mathrm{max}}$ and susceptibility $\chi^{\mathrm{max}}$, 
as well as minimum of the fourth order cumulant $U_4^{\mathrm{min}}$ as function of $L^2$.\label{schulz14}}
\end{figure*}
%
%%%%%%%%%%%%%%%%%%%%%%%%%%%%%%%%%%%%%%%%%%%%%%%%%%%%%%%%%%%%%%%%%%%%%%%%%%%%%%%%%%%%%%%%%%%%%%%%%%%%%%%%%%%%%%%%%%
%
%
A similar argumentation applies to the distribution of the order parameter $P_{L,D}(m)$ \cite{Vollmayr} yielding the same scaling 
behavior for the susceptibility $\chi$, i.e.
\begin{eqnarray}
\beta_{\chi^{\mathrm{max}}}(D,L)-\beta_{\mathrm{tr}}(D) & \propto & (L^2D)^{-1},\\
\chi^{\mathrm{max}} & \propto &  L^2D, \label{chimax}
\end{eqnarray}
and one can show that the fourth order cumulant $U_4$ (Fig.~\ref{schulz12}) takes a minimum value at an inverse temperature
$\beta_{U_4^{\mathrm{min}}}(D,L)$ which is again shifted 
\begin{equation}
\beta_{U_4^{\mathrm{min}}}(D,L)-\beta_{\mathrm{tr}}(D)\propto (L^2D)^{-1}, \label{shiftTCU} 
\end{equation}
while the minimum $U^{\mathrm{min}}_4$ obeys 
\begin{eqnarray}
U^{\mathrm{min}}_4 & \propto & - L^2D.\label{cumax}
\end{eqnarray}
Furthermore it was shown \cite{Vollmayr}, that the shift in the crossing points of the cumulants for different system sizes is proportional to
$N^{-2}$, which is negligibly small on the scale of $N=L^2 D$.\\
Fig.~\ref{schulz14}(b) now shows the maximum values of the response functions $c^{\mathrm{max}}$, $\chi^{\mathrm{max}}$, and the minimum 
$U_4^{\mathrm{min}}$ of the cumulant as function of $1/L^2$ for the three different thicknesses $D=6,8,12$. As can be seen from the plots, 
the data comply well with the behavior predicted by expressions (\ref{specmax}),(\ref{chimax}) and (\ref{cumax}). 
Considering the fourth order cumulant $U_4$ in case of $D=6$, one observes that sub-leading corrections to scaling are still present 
for the smaller linear dimensions $L$, but the expected linear behavior in $L^2$ is born out for the three largest choices of $L$.\\
The definition for the finite lattice transition temperature considered so far, e.g. Eq.~(\ref{betashift}), involve 
leading order corrections of $1/L^2$. 
An alternative definition of the transition temperature which has the additional benefit that the latter corrections are absent was given in 
Ref.~\cite{JankeBorgs}. Here, it is utilized that at the infinite-lattice transition point $\beta_{\mathrm{tr}}(D)$ all phases coexist 
which implies that the sum of the weights of the $q$ ordered phases equals $q$ times the weight of the disordered phase, i.e., 
\begin{equation}
R(\beta_{\mathrm{ew}},L,D)\equiv\frac{\sum_{e\le e_{\mathrm{cut}}} P_{L,D}(e,\beta_{\mathrm{ew}})  }{\sum_{e>e_{\mathrm{cut}}} 
P_{L,D}(e,\beta_{\mathrm{ew}})  }=q ,\label{Rcond}
\end{equation}
where $P_{L,D}(e)$ is the (finite-size) energy probability distribution, and $\beta_{\mathrm{ew}}(D,L)$ differs from $\beta_{\mathrm{tr}}(D)$ 
only by corrections exponentially small in system size. The energy $e_{\mathrm{cut}}$ appearing in Eq.~(\ref{Rcond}) is taken to be the 
internal energy at the temperature where the specific heat is maximal \cite{JankeBorgs}. 
\subsection{Transition temperatures}
\begin{table}[t]
\begin{center}
\begin{tabular}{clllll} \hline\hline
$D$ & $\Delta e(D)/J$ &$J\beta_{c^{\mathrm{max}}}(D)$ &$J\beta_{U_4^{\mathrm{min}}}(D)$ & $J\beta_{\mathrm{ew}}(D)$ & 
$J\beta_{\mathrm{tr}}(D)$ 
\\ \hline 
6 &0.261(6)&0.24082(2)&0.24079(4)&0.24082(1) & $0.24082(4)$\\  
8 &0.300(2)&0.24726(3)&0.24716(7)&0.24725(2) & $0.24725(10)$\\   
12 &0.236(3)&0.25115(4)&0.25109(4)&0.25117(2)& $0.25115(10)$\\ \hline 
\end{tabular}
\end{center}
\caption[Estimates of the latent heats $\Delta e(D)$ and the inverse transition temperatures $\beta_{\mathrm{tr}}(D)$ (thin Ising film, 
$J_{\mathrm{s}}/J=1.5$).]{Estimates for the latent heats $\Delta e(D)$ and the inverse transition temperatures of the first order interface 
localization-delocalization transition for different film thicknesses $D$. $\beta_{c^{\mathrm{max}}}(D,\infty)$ and 
$\beta_{U_4^{\mathrm{min}}}(D,\infty)$ are the estimates of the transition point $\beta_{\mathrm{tr}}(D)$ originating from an extrapolation 
of peak positions as described in the text, while $\beta_{ew}(D,\infty)$ denotes the estimate from the equal weight rule (\ref{Rcond}). The 
final estimate of the inverse temperature $\beta_{\mathrm{tr}}(D)$ of the triple point is stated in the last column.
\label{estimatesofpoint}}
\end{table}
Now, we can extract the infinite volume transition point $\beta_{\mathrm{tr}}(D)$ from the finite size data, i.e., as Eqs.~(\ref{betashift}) 
and (\ref{shiftTCU}) suggests by
fitting the peak positions for fixed $D$ to
\begin{equation}
\beta_{\mathrm{max,min}}(D,L)=\beta_{\mathrm{max,min}}(D,\infty)+\frac{a}{L^2}, \label{fitdata}
\end{equation}  
where $\beta_{\mathrm{max,min}}(D,L)$ stands for the location of the maximum of the specific heat $\beta_{c^{\mathrm{max}}}(D,L)$ 
and the location of the minimum $\beta_{U_4^{\mathrm{min}}}(D,L)$ of the fourth order cumulant at finite $L$, while $\beta_{\mathrm{max,min}}(D,\infty)$
denotes the infinite volume limit ($L\rightarrow \infty$) of the corresponding inverse temperatures, which is an estimate of the infinite 
system transition point $\beta_{\mathrm{tr}}(D)$. 
Alternatively, we have also 
employed the finite volume estimator $\beta_{\mathrm{ew}}(D,L)$ of the transition point, 
as defined by the condition (\ref{Rcond}).\\
The individual results for the infinite system transition points 
%
%and the best estimates for the latent heats 
%
are summarized in Table {\ref{estimatesofpoint}. In the last column of Table \ref{estimatesofpoint} we state our final estimate of the infinite 
system transition point $\beta_{\mathrm{tr}}(D)$, based on weighted averages over the estimates listed in column 2--4.
Concerning the error in our final estimate of $\beta_{\mathrm{tr}}(D)$ we have also accounted for the scatter in the crossings of the energy curves
as depicted in Fig.~\ref{schulz9} and the crossings in the fourth order cumulant $U_4$, see Fig.~\ref{schulz12}. 
While we find that the order of magnitude of the error as determined from the various finite lattice estimators considered above, 
complies well with all the data for $D=6$, especially the latter crossing points, 
we may have uncontrolled errors in case of the larger thicknesses $D=8$ and $D=12$, due to the aforementioned lack of statistics 
deep in the pure phases. In these cases we consider here as a conservative error estimate the extremal crossing points as an upper bound
to the transition point, which results in the error of $\beta_{\mathrm{tr}}(D\ge8)$ as given in the last column of Table~\ref{estimatesofpoint}.
Fitting the locations of the maxima of the specific heat to Eq.~(\ref{fitdata}), as depicted in Fig.~\ref{schulz14}(a), 
one can also determine the latent heat $\Delta e$ which is however less
accurate than computing $\Delta e$ from the distribution $P_{L,D}(E)$ via \cite{Billoire}
\begin{equation}
\Delta e(L,D) = \Delta e(D)+\mathrm{const}\times L^{-2}, \label{billext}
\end{equation}
which yields the values stated in column $2$ of Table~\ref{estimatesofpoint}.
Concerning the extrapolation (\ref{fitdata}) of the positions of the minima $U_4^{\mathrm{min}}$ and the maxima $c^{\mathrm{max}}$ we 
have used only data for $L>32$ in case of $D=6$. For these lattices, exponential corrections to $\beta_{\mathrm{ew}}(D,L)$ cannot be 
resolved within the achieved accuracy. This is also the case for the larger film thicknesses $D$ and all choices of $L$. 
Hence, the values listed in Table~\ref{estimatesofpoint} for $\beta_{\mathrm{ew}}(D)$ are simply averages over the various 
lateral system sizes $L$ ($L>32$ in case of $D=6$). 
%
%
%%%%%%%%%%%%%%%%%%%%%%%%%%%%%%%%%%%%%%%%%%%%%
%%%%%%%%%%%%%%%%%%%%%%%%%%%%%%%%%%%%%%%%%%%%%%
%%%%%%%%%%%%%%%%%%%%%%%%%%%%%%%%%%%%%%%%%%%%%
%
\subsection{\label{wett} Wetting Temperature of the semi-infinite system}
%
%%%%%%%%%%%%%%%%%%%%%%%%%%%%%%%%%%%%%%%%%
%%%%%%%%%%%%%%%%%%%%%%%%%%%%%%%%%%%%%%%%%
%%%%%%%%%%%%%%%%%%%%%%%%%%%%%%%%%%%%%%%%%%
%
%%%%%%%%%%%%%%%%%%%%%%%%%%%%%%%%%%%%%%%%%%%%%%%%% schulz15 %%%%%%%%%%%%%%%%%%%%%%%%%%%%%%%%%%%%%%%%%%%%%%%%%%%%%%%
%
\begin{figure}[b]
\includegraphics[clip, width=0.46\textwidth]{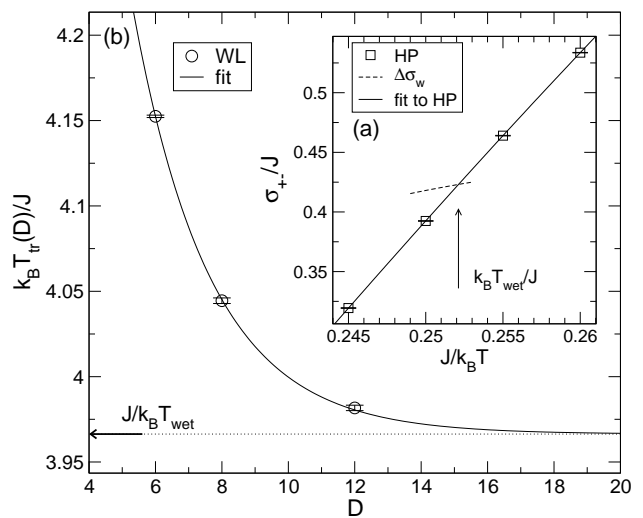}
\caption{(a) Shown are the interfacial tension $\sigma$ of the $3D$ Ising-model (HP) taken from Ref.~\cite{HP}, 
fitted by an $8^{\mathrm{th}}$ degree polynomial in order to smoothly interpolate between the data points as well as the 
the quantity $\Delta \sigma_{\mathrm{w}}$ appearing in the Young equation (\ref{young}). 
The position of the crossing point yields the wetting temperature 
$J\beta_{\mathrm{w}}(H_1)=J/k_{\mathrm{B}}T_{\mathrm{w}}(H_1)=0.25212(5)$. \label{schulz15}}
\end{figure}
%
%%%%%%%%%%%%%%%%%%%%%%%%%%%%%%%%%%%%%%%%%%%%%%%%%%%%%%%%%%%%%%%%%%%%%%%%%%%%%%%%%%%%%%%%%%%%%%%%%%%%%%%%%
%
In order to determine the wetting temperature  \mbox{$\beta_{\mathrm{w}}=\lim_{D\rightarrow \infty} \beta_{\mathrm{tr}}(D)$} of the semi-infinite 
system, we have studied Hamiltonian (\ref{thini}) with $D=12$ and $L=48$ 
along the branch of positive bulk-magnetization at the inverse temperature $\beta=0.251$ near the expected location of the 
wetting temperature $\beta_{\mathrm{w}}(H_1)$. 
We have performed simulations for five different 
sets of surface fields (symmetric, i.e., $H_1=H_D$), namely $H_1/J=-0.25$, $-0.125$, $0$, $0.125$, and $0.25$, utilizing a conventional 
Metropolis algorithm in order to measure the surface magnetization \mbox{$\langle m_1  \rangle=\langle \sum_{i \in \mathrm{surface} 1} S_i \rangle/N$} 
using up to $10^7$ MCS for averaging. This selection of surface fields allows one to reweight to all fields in the range $[-0.25J,0.25J]$ for a range of 
inverse temperatures $J\beta \in [0.249,0.253]$. 
Note, that the metastability is strong enough (cf. Fig.~\ref{schulz3}) that the system remains in the ordered phase (initially all spins up) 
even for $H_1/J=-0.25$.
According to the Young equation \cite{Young} the walls are wet by spin down, if the difference $\Delta \sigma_{\mathrm{w}}$ 
between the surface free energy of the wall with respect to a positively magnetized bulk $\sigma_{\mathrm{w}+}$ and the surface free energy against 
a negatively magnetized bulk $\sigma_{\mathrm{w}-}$ exceeds the interfacial tension $\sigma$ of the 3D Ising-model \cite{HP} at an infinite 
distance from the wall.
\begin{equation}
\Delta \sigma_{\mathrm{w}} = \sigma_{\mathrm{w}+}-\sigma_{\mathrm{w}-} > \sigma \label{young}
\end{equation}
By symmetry $\sigma_{\mathrm{w}-}(-H_1)$ equals $\sigma_{\mathrm{w}+}(H_1)$, i.e., the free energy cost of a wall favoring 
spin up with respect to a positively magnetized bulk. Thus we can perform a thermodynamic integration \cite{WET}
\begin{eqnarray}
\Delta \sigma_{\mathrm{w}}& =& \sigma_{\mathrm{w}+}(-H_1)-\sigma_{\mathrm{w}+}(H_1) \nonumber \\
& = & \int_{-H_1}^{H_1} \mathrm{d}H_1' \langle m_1(H_1')\rangle_{\beta},\quad H_1=0.25J
\end{eqnarray}
and determine the wetting temperature $\beta_{\mathrm{w}}(H_1)$ by the condition \mbox{$\Delta \sigma_w=\sigma$},
which yields \mbox{$J\beta_{\mathrm{w}}(H_1)=0.25212(5)$} as depicted in Fig.~\ref{schulz15}(a).\\
Describing the semi-infinite system by means of the wetting film 
thickness $l$ leads to the effective interface potential \cite{ParryEvans}
\begin{equation}
V_{\mathrm{eff}}(l)= a\exp(-\kappa l) -b\exp(-2\kappa l) +c\exp(-3\kappa l), \label{preciseform}
\end{equation}
which has the meaning of a free energy cost when placing a (flat) interface at distance $l$ from the wall. Upon minimizing $V_{\mathrm{eff}}(l)$ 
with respect to $l$ one finds the equilibrium position of the interface. Eq.~(\ref{preciseform}) includes only the lowest powers of 
$\exp(-\kappa l)$ which are necessary to describe a first order wetting transition in the semi-infinite system. The coefficient $a$ 
explicitly depends on 
temperature, while the temperature dependence of $b$ and $c$ is neglected ($c>0$ in the following) \footnote{The description of the interface 
in terms of the effective interface potential $V_{\mathrm{eff}}$ follows from the sharp-kink approximation to the capillary wave Hamiltonian 
\mbox{$\mathcal{H}_{\mathrm{eff}}=\int \mathrm{d}\mathbf{\rho}[(\sigma/2)(\nabla l)^2+V_{\mathrm{eff}}\{l(\mathbf{\rho}  \}]$} where 
fluctuations of the local interface position are neglected.}. 
All coefficients have the same magnitude as the 
interfacial tension between bulk phases and one finds a first order wetting transition for $b>0$ at $a_{\mathrm{w}}=b^2/4c$, 
where the interface jumps discontinuously into the bulk \cite{MuellerBinderAlbano,MuellerAlbanoBinder}. 
Now, for a film one has an additional contribution from the second wall and the effective potential reads \cite{MuellerBinderSymPol}
\begin{eqnarray}
\Delta V_{\mathrm{eff,Film}}(l)&=& V_{\mathrm{eff}}(l)+V_{\mathrm{eff}}(D-l)-2V_{\mathrm{eff}}(D/2)\nonumber \\
                               &=& c\left[\tilde{m}^2(\tilde{m}^2-r)^2+t\tilde{m}^2\right],\label{pot}
\end{eqnarray}
with
\begin{equation}
r=\frac{b-6c\exp(-\kappa D/2)}{2c},
\end{equation}
and
\begin{equation}
t=\frac{a-a_{\mathrm{w}}-b\exp(-\kappa D/2)}{c}. \label{fff}
\end{equation}
In Eq.~(\ref{pot}) we have utilized the auxiliary variable 
\begin{eqnarray}
\tilde{m}&=&2\exp(-\kappa D/2)\{\cosh \left[ \kappa\left(l-D/2\right)\right]-1\} \nonumber \\
         &=&(\exp(-\kappa D/4)\kappa [l-D/2])^2 \nonumber \\
         & &  +\mathrm{higher\;orders\;of\;} [l-D/2].
\end{eqnarray}
In the film, $r>0$ gives rise to first order interface localization-delocalization transitions and $t=0$ then denotes 
the triple temperature. Hence, for large $D$ we have from Eq.~(\ref{fff})
\begin{equation}
a_{\mathrm{tr}}=a_{\mathrm{wet}}+ b \exp(-\kappa D/2),\label{expfit}
\end{equation}
i.e., the triple temperature differs from the wetting temperature only by a term exponentially small in $\kappa D/2$ and 
is larger than the wetting temperature ($b>0$). 
Within mean field theory $\kappa$ would have to be identified with the inverse bulk correlation length $\xi_{\mathrm{b}}$ \cite{ParryEvans}. 
However, from the two-field Hamiltonian approach developed in Ref.~\cite{BoulterParry} we know that $\kappa/2$ has to be replaced by 
\begin{equation}
\frac{\kappa}{2}=\frac{1}{2 \xi_{\mathrm{b}}\theta},\quad \theta=1+\omega_{\mathrm{eff}}/2, \label{kkk}
\end{equation}
where $\omega_{\mathrm{eff}}$ is the effective wetting parameter which becomes $\lim_{T\rightarrow T^+_{\mathrm{w}}}\omega_{\mathrm{eff}}=
k_{\mathrm{B}}T/4\pi\sigma\xi_{\mathrm{b}}^2$ upon lowering the temperature $T$ towards the wetting temperature $T_{\mathrm{w}}$ \cite{Swain}. 
From a simple exponential fit of the form 
(\ref{expfit}) we get $\kappa/2=0.430(8)$. (Note, that this has to be regarded as an effective value since we neglect any temperature 
dependence of $\kappa$ within our range of triple temperatures $\beta_{\mathrm{tr}}(D)$). Evaluating now $\theta$ at $T_{\mathrm{w}}/T_{cb}=0.88$ 
where we employ $\xi_{\mathrm{b}}\sim 0.88$ \cite{HP}, yields $\theta \sim 1.3$, 
which is compatible with the values extracted for $\theta$ by Parry {\itshape et al.} \cite{Swain} 
%(and the estimate $k_{\mathrm{B}}T/4\pi\sigma\xi_{\mathrm{b}}^2 \approx 0.64(6)$) 
and clearly differs from the value $\theta=1$ 
expected from mean-field theory. 
Of course, making more quantitative statements would require data from additional 
film thicknesses $D$, but the above considerations clearly indicate that our data nicely supports the asserted functional dependence
of $\beta_{\mathrm{tr}}(D)$ on $D$, i.e., Eq.~(\ref{expfit}).
%
%
%%%%%%%%%%%%%%%%%%%%%%%%%%%%%%%%%%%%%%%%%%%%%
%%%%%%%%%%%%%%%%%%%%%%%%%%%%%%%%%%%%%%%%%%%%%%
%%%%%%%%%%%%%%%%%%%%%%%%%%%%%%%%%%%%%%%%%%%%%
%
\section{\label{concl} Conclusion}
%
%%%%%%%%%%%%%%%%%%%%%%%%%%%%%%%%%%%%%%%%%
%%%%%%%%%%%%%%%%%%%%%%%%%%%%%%%%%%%%%%%%%
%%%%%%%%%%%%%%%%%%%%%%%%%%%%%%%%%%%%%%%%%%
%
We have studied the interface localization-delocalization transition in a thin Ising-film (\ref{thini}) for a choice of parameters, 
where the transition is pronounced first order for all studied thicknesses $D=6$, $8$, and $12$.
Checking for the correct behavior of the logarithm of the partition function $\ln Z$ which should converge to $N\ln2$ as $\beta\rightarrow 0$,
we find reasonable agreement for $D=6$ within error bars (cf. Table \ref{freeenergylargeT}) 
In contrast, for $D>6$ we see rather clear deviations from the expected value with relative deviations up to $10^{-3}$. 
We attribute this behavior to a slowing down encountered in the flat energy-histogram ensemble. 
Difficulties also arise, when one considers to sample a flat magnetization distribution, 
although simulation results suggest that the slowing down is less severe. Here, we find evidence for a discontinuous shape transition, 
as studied by Neuhaus and Hager \cite{NeuhausHager}. 
For the larger thicknesses ($D>6$) we therefore suggest to employ an additional reference for the disordered phase (total number of states), 
in order to get the proper relative weight between the coexisting phases, thus correcting for the lack of tunneling events, 
in the late stages of the algorithm. 
The triple temperatures $\beta_{\mathrm{tr}}(D)$ of the interface localization-delocalization transition can then be determined
with a relative accuracy of the order $10^{-4}$ while the relative error in the latent heats is of the order $10^{-2}$.
The triple temperatures are seen to differ from the wetting temperature of the semi-infinite system 
by a term exponentially small in film thickness $D$ as predicted by the 
sharp-kink approximation to the capillary wave Hamiltonian, provided the length scale $\kappa$ is identified with the results of 
Parry and co-workers, i.e., Eq.~(\ref{kkk}).\\
When one compares the present results based on Wang-Landau sampling \cite{WangLandau1,WangLandau2,SchulzBinderMueller,SchulzBinderMuellerLandau} 
to the first study of first order interface localization-delocalization transitions \cite{Ferrenberg} where simple Metropolis and heatbath Monte 
Carlo algorithms were used, a major improvement of accuracy is clearly seen. On the other hand, the systematic problems due to entropic barriers 
described in our work show that it would be problematic to apply the Wang-Landau algorithm to larger systems than used here. Note, that the 
largest sizes used by us, $128 \times 128 \times 12\sim 1.97\cdot 10^5$ Ising spins, distinctly exceed the sizes analyzed in most previous applications 
of this algorithm \cite{WangLandau1,WangLandau2,SchulzBinderMueller,SchulzBinderMuellerLandau}.
\begin{acknowledgments}
This work was supported in part by the Deutsche Forschungsgemeinschaft under grants No Bi314/17 and Tr6/c4. Helpful and stimulating 
discussions with D. P. Landau and P. Virnau are gratefully acknowledged. We thank NIC J\"ulich and HLR Stuttgart for a grant of computer time at 
the CRAY-T3E supercomputer.
\end{acknowledgments}
\bibliography{schulz}
\end{document}